\documentclass[prb,twocolumn,superscriptaddress,showpacs]{revtex4-1}

\pdfoutput=1

\usepackage{graphicx}% Include figure files
\usepackage{amsmath,amssymb}

\usepackage{ulem}
\usepackage{color}
\usepackage{amssymb}
\usepackage{braket}

\def\diff{\mathrm d}
\def\mathi{\mathrm i}

\def\diff{\mathrm d}
\def\mathi{\mathrm i}

\newcommand{\mcave}[1]{\ensuremath{\braket{#1}_\mathrm{MC}}}

\newcommand{\opcdag}{\ensuremath{c^\dagger}}
\newcommand{\opc}{\ensuremath{c}}

\begin{document}
\title{
Negative sign problem in continuous-time quantum Monte Carlo: optimal choice of single-particle basis for impurity problems
}
\author{Hiroshi Shinaoka} 
\affiliation{Theoretische Physik, ETH Zurich, 8093 Zurich, Switzerland}
\affiliation{Department of Physics, University of Fribourg, 1700 Fribourg, Switzerland}
\author{Yusuke Nomura}
\affiliation{Centre de Physique Th\'{e}orique, \'{E}cole Polytechnique, CNRS UMR7644, 91128 Palaiseau Cedex, France}
\author{Silke Biermann}
\affiliation{Centre de Physique Th\'{e}orique, \'{E}cole Polytechnique, CNRS UMR7644, 91128 Palaiseau Cedex, France}
\author{Matthias Troyer}
\affiliation{Theoretische Physik, ETH Zurich, 8093 Zurich, Switzerland}
\author{Philipp Werner}
\affiliation{Department of Physics, University of Fribourg, 1700 Fribourg, Switzerland}
\date{\today}

%%%-----------------------------------------------------------------
\begin{abstract}
The negative sign problem in quantum Monte Carlo (QMC) simulations of cluster impurity problems is the major bottleneck in cluster dynamical mean field calculations.
In this paper we systematically investigate the dependence of the sign problem on the single-particle basis. We explore both the hybridization-expansion and the interaction-expansion variants of continuous-time 
QMC for three-site and four-site impurity models with baths that are  diagonal in the orbital degrees of freedom.
We find that the sign problem in these models can 
be substantially reduced by using a non-trivial single-particle basis. 
Such bases can be generated by diagonalizing a subset of the intracluster hoppings.
\end{abstract}

% 71.10.Fd: Lattice fermion models (Hubbard model, etc.)

\maketitle

\section{Introduction}
Quantum Monte Carlo (QMC) methods are a powerful tool for studying the properties of quantum many-body systems. They are based on a mapping of the quantum system to an auxiliary classical system. which is then sampled stochastically. 
A fundamental limitation of fermionic QMC algorithms is the so-called negative sign problem,~\cite{LohJr:1990up}
which appears when configurations with negative weights appear due to fermionic statistics.
Apart from models with special symmetries,~\cite{Hirsch:1985wz,Chandrasekharan:1999kn,Capponi:2001jc,Wu:2003es,Wu:2005im,Li:2015cw,Li:2015jf,Wang:2015vha}
the sign problem prevents access to the low-temperature bulk properties of many-body fermionic systems since the efficiency of the MC sampling decreases exponentially with  system size and inverse temperature as detailed below.

In a QMC algorithm, we express the partition function of the quantum system as the sum of 
contributions of classical configurations $\{c\}$, which define a configuration space $\Omega$,
\begin{eqnarray}
Z &=& \mathrm{Tr}[e^{-\beta H}] = \sum_{c\in \Omega} w_c.
\end{eqnarray}
The expectation value of an observable $\hat{O}$ is then given by
\begin{eqnarray}
\braket{\hat{O}} &=& \frac{\mathrm{Tr}[\hat{O} e^{-\beta H}] }{\mathrm{Tr}[e^{-\beta H}] } = Z^{-1} \sum_{c\in \Omega} O_c w_c,
\end{eqnarray}
where $O_c$ is the value of the observable associated with $c$.
When $w_c\ge 0$ ($\forall c$), $\braket{\hat{O}}$ can be efficiently evaluated by a summation over $N_c$ configurations sampled from $\Omega$ according to the weight $\{w_c\}$, 
\begin{eqnarray}
\braket{\hat{O}} &\simeq& \braket{\hat{O}}_\mathrm{MC} \equiv \frac{\sum_i O_{c_i}}{N_c}.
\end{eqnarray}
The standard deviation $\delta \braket{\hat{O}}$ of this MC estimate scales as $1/\sqrt{N_c}$ when $N_c$ is larger than the autocorrelation time of the MC dynamics.

The MC sampling suffers from a negative sign problem when $w_c< 0$ for some configurations.
In this case, one cannot interpret $w_c/Z$ as the probability for the configuration $c$. 
Using $|w_c|$ for the weight instead,
one can still perform importance sampling as follows:
\begin{eqnarray}
\braket{\hat{O}} &=& \frac{\sum_c \mathrm{sign}(c)~ O(c)}{\sum_c \mathrm{sign}(c)} = \frac{\mcave{\mathrm{sign} ~\hat{O}}}{\mcave{\mathrm{sign}}}.\label{eq:signed-obs}
\end{eqnarray}
However, in simulations with a sign problem, both $\mcave{\mathrm{sign} ~\hat{O}}$ and $\mcave{\mathrm{sign}}$ decay exponentially with increasing system size and inverse temperature. 
This makes the MC sampling exponentially inefficient.
To keep the error on $\langle \hat O \rangle$ constant, 
we need to increase the number of Monte Carlo steps (at least) as $1/\mcave{\text{sign}}^2$.

In this paper we investigate the fermionic sign problem for so-called quantum impurity models.~\cite{Ferrero:2007bz}
These describe a small number of interacting orbitals hybridized with a noninteracting bath.
The simplest of these models is the single-orbital Anderson impurity model which was originally proposed to describe a magnetic impurity embedded in a metal.~\cite{Anderson:1961,Hewson:1997vc}
Subsequently, it was found that in infinite dimensions, 
the Hubbard model can be exactly mapped onto a single-site impurity model
with a bath which is self-consistently determined.~\cite{VanDongen:1990df,Metzner:1989bz}
In finite dimensions, the analogous procedure leads to the dynamical mean-field theory (DMFT) approximation for correlated lattice models.~\cite{Georges:1996un}
DMFT can be extended to incorporate short-range correlations~\cite{Maier:2005et} by mapping onto cluster-type impurity systems, as done in schemes such as cluster DMFT,~\cite{Lichtenstein:2000bp, Kotliar:2001iy}
and the dynamical cluster approximation.~\cite{Hettler:2000es}
The success of DMFT created a demand for powerful and versatile impurity solvers and led to the development of continuous-time Monte Carlo impurity solvers.  

There are two complementary types of continuous-time quantum Monte Carlo algorithms for solving an impurity problem, which are both based on a stochastic sampling of a perturbation expansion:
the interaction expansion method (CT-INT),~\cite{Rubtsov:2005iwa} 
the auxiliary-field Monte Carlo method,~\cite{Gull:2008cma}
and the hybridization expansion method (CT-HYB).~\cite{Werner:2006ko,Werner:2006iz}
These methods do not have a sign problem at any filling in the case of the single-orbital impurity model.~\cite{Rubtsov:2003vb,Yoo:2005ib,Werner:2006ko}
However, they generally suffer from a negative sign problem when applied to a multi-orbital or cluster impurity model with an internal structure in which electrons can be exchanged and the sign problem becomes worse as the system becomes larger.

Since the sign problem is not gauge invariant and representation dependent, the severity of the sign problem depends on the single-particle basis used to represent the impurity.
For CT-HYB, it has been empirically known that the sign problem is sometimes improved by using the single-particle basis that diagonalizes the intracluster single-particle Hamiltonian of the impurity.~\cite{Furukawa:2010bx,Anonymous:5bgqkPSw}
However, to the best of our knowledge, no systematic effort has yet been made to clarify how the sign problem depends on the single-particle basis for CT-HYB.
On the other hand, for CT-INT, the site basis, with local interactions identical to those of the lattice model,
is commonly used in solving Hubbard-like cluster impurity models.
There, one faces a severe sign problem particularly in systems away from half filling and in geometrically frustrated systems.
However, there has been no previous effort to test alternative single-particle bases to reduce the sign problem.

In this work, we systematically investigate the single-particle-basis dependence of the sign problem for typical cluster impurity models within CT-HYB and CT-INT.
We focus on a trimer model and a tetramer impurity model with diagonal baths.
The main conclusion of this study is that the sign problem can be dramatically improved by choosing a single particle basis which diagonalizes some part of the intracluster single-particle Hamiltonian.

The rest of this paper is organized as follows.
Section II is devoted to the results obtained for CT-HYB.
Section II A describes the formalism and the implementation of the basis rotation. 
We present the results for the trimer and the tetramer models in Secs.~II B and II C, respectively.
Section~III discussed the basis dependence for CT-INT.
The implementation is described in Sec.~III A.
Then, we show the results of the trimer and tetramer models in Secs.~III B and III C, respectively. 
A summary and discussion are provided in Sec.~IV.

\section{Hybridization-expansion} % continuous-time Monte Carlo method}
\subsection{Implementation}

In this section,we review the hybridization expansion continuous-time quantum Monte Carlo algorithm (CT-HYB) for quantum impurity problems.~\cite{Werner:2006ko,Werner:2006iz}
Tracing out the bath degrees of freedom,
the effective action is given by
\begin{eqnarray}
  S &=& S_\mathrm{loc} + \sum_{ab} \int_0^\beta  \diff \tau \diff \tau^\prime \Delta_{ab}(\tau^\prime-\tau) c^\dagger_a(\tau^\prime) c_b(\tau),\hspace{5mm}\label{eq:action-cthyb}
\end{eqnarray}
where $\Delta$ is the hybridization function [$\Delta_{ab}(\tau) = \Delta_{ba}^*(\tau)$] and 
$a$, $b$ are combined spin and site indices.
We denote the inverse temperature by $\beta$. 
If $S_\mathrm{loc}$ contains only instantaneous terms up to two-body interactions, the corresponding Hamiltonian reads
\begin{eqnarray}
  \mathcal{H}_\mathrm{loc} &=& -\sum_{ab} t_{ab} c^\dagger_a c_b + \sum_{abcd} V_{abcd}c_a^\dagger c_b^\dagger c_c c_d.
   \label{eq:Simp}
\end{eqnarray}

We may define a transformed single-particle basis 
\begin{eqnarray}
  d_a &=& \sum_b U_{ba}^* c_b,\\
  d^\dagger_a &=& \sum_b U_{ba} c^\dagger_b,
\end{eqnarray}
with $U_{ab}$ being a unitary matrix. In this basis the second term in Eq.~(\ref{eq:action-cthyb}) reads
\begin{eqnarray}
  S_\mathrm{hyb} &=& \sum_{ab} \int_0^\beta  \diff \tau \diff \tau^\prime \bar{\Delta}_{ab}(\tau^\prime-\tau) d^\dagger_a(\tau^\prime) d_b(\tau),\hspace{3mm}\label{eq:action-cthyb-rot}
\end{eqnarray}
with
\begin{eqnarray}
\bar{\Delta}_{ab} (\tau) &=& \sum_{cd} (U^\dagger)_{ac} \Delta_{cd}(\tau) U_{db}.
\end{eqnarray}

Now, we expand the partition function $Z$ as
\begin{eqnarray}
  Z &=& Z_\mathrm{bath}  \sum_{n=0}^\infty \frac{1}{n!^2}\sum_{\alpha_1,\cdots,\alpha_n}\sum_{\alpha^\prime_1,\cdots,\alpha^\prime_n}\nonumber \\
  &&  \int_0^\beta \mathrm{d} \tau_1 \mathrm{d} \tau_1^\prime \cdots \int_0^\beta \mathrm{d} \tau_n \mathrm{d} \tau_n^\prime \nonumber\\
   && \mathrm{Tr_{loc}}\left[ e^{-\beta\mathcal{H}_\mathrm{loc}} T d_{\alpha_n}(\tau_n) d^\dagger_{\alpha_n^\prime}(\tau_n^\prime) \cdots d_{\alpha_1}(\tau_1) d^\dagger_{\alpha_1^\prime}(\tau_1^\prime)\right]\nonumber\\
   && \times \mathrm{det} \boldsymbol{\bar{M}}^{-1},\label{eq:z-cthyb}
\end{eqnarray}
where $Z_\mathrm{bath}$ is the partition function of the bath.
The matrix elements of $(\boldsymbol{\bar{M}}^{-1})_{ij}$ are given by the hybridization function $\bar{\Delta}_{\alpha_i^\prime,\alpha_j}(\tau_i^\prime - \tau_j)$.

The partition function in Eq.~(\ref{eq:z-cthyb}) can be evaluated by importance sampling of configurations of annihilation and creation operators on the imaginary-time axis.
The weights are given by % CHANGED d\tau^{n} TO d\tau^{2n}
\begin{align}
  &w(d_{\alpha_1}(\tau_1), \cdots, d_{\alpha_n}(\tau_n); d_{\alpha_1^\prime}(\tau_1^\prime), \cdots, 
  d_{\alpha_1^\prime}(\tau_n^\prime)) \nonumber \\
  &\hspace{10mm}= \frac{d\tau^{2n}}{n!^2} \mathrm{Tr_{loc}}\Big[ e^{-\beta\mathcal{H}_\mathrm{loc}} T d_{\alpha_n}(\tau_n) d^\dagger_{\alpha_n^\prime}(\tau_n^\prime) \cdots \nonumber \\  
  &\hspace{32mm} d_{\alpha_1}(\tau_1) d^\dagger_{\alpha_1^\prime}(\tau_1^\prime)\Big] \mathrm{det} \boldsymbol{\bar{M}}^{-1}.\label{eq:weight}
\end{align}
The local trace is evaluated using the matrix formalism
where the operators $e^{-\tau\mathcal{H}_\mathrm{loc}}$, $d_\alpha$, $d_\alpha^\dagger$ in Eq.~(\ref{eq:weight}) are represented in the eigenbasis of $\mathcal{H}_\mathrm{loc}$.~\cite{Werner:2006iz}
Note that we do not have to transform $\mathcal{H}_\mathrm{loc}$ to the new single-particle basis.

\subsection{Trimer model}
\subsubsection{Set-up}
In this section, we study a trimer impurity model as a minimal system which exhibits a negative sign problem.
As shown in Fig.~\ref{fig:trimer},
each site is connected to a bath with a semicircular density of states of width $4$.
The Hamiltonian is given by
\begin{eqnarray}
 \mathcal{H} &=& -\sum_{i\neq j}^3 t_{ij} \opcdag_{i\sigma} \opc_{j\sigma} + U\sum_{i=1}^3 \hat{n}_{i\uparrow} \hat{n}_{i\downarrow}-\mu\sum_{i=1}^3 \hat{n}_i\nonumber\\
  && + \lambda \sum_{i=\alpha}\sum_{k\sigma} (\opcdag_{i\sigma} a_{\alpha k \sigma} +a^\dagger_{\alpha k \sigma} \opc_{i\sigma} )\nonumber\\
  && + \sum_{\alpha} \sum_{k \sigma} \epsilon_k a^\dagger_{\alpha k \sigma} a_{\alpha k \sigma}
,\label{eq:trimer}
\end{eqnarray}
with $\mu=U/2$ and $\lambda=1$.
Throughout this section, we measure energy in units of $\lambda$.
We set the onsite Coulomb repulsion to $U=5$ throughout this section.
The intracluster hopping matrix elements are given by $t_{12}=t^\prime$, $t_{13}=t$, $t_{23}=t^\prime$ ($t_{ij}=t_{ji}$) [see Fig.~\ref{fig:trimer}(a)].
$\alpha$ is the index of the bath and the different bath levels are labeled by $k$.
The distribution of their energy levels $\epsilon_k$ is given by $\sum_k \delta (\epsilon - \epsilon_k) = \frac{1}{2 \pi} \sqrt{4 - \epsilon^2}$, that is, the hybridization function $\Delta$ is given by
\begin{eqnarray}
\Delta_{ab}(\tau)&=& \frac{\delta_{ab}}{\beta} \sum_{n=-\infty}^\infty e^{-\mathrm{i}\omega_n \tau} \int_{-2}^2 d \epsilon \frac{\frac{1}{2\pi}\sqrt{4-\epsilon^2}}{\mathi \omega_n - \epsilon}\nonumber\\
&=&\frac{\delta_{ab}}{\beta}  \sum_{n=0}^\infty (\omega_n-\sqrt{\omega_n^2+4})\sin({\omega_n \tau})~(0<\tau<\beta),\nonumber\\
\end{eqnarray}
where $\omega_n = (2n+1)\pi/\beta$ is the Matsubara frequency.~\cite{Georges:1996un}
Note that the hybridization function is invariant with respect to single-particle basis rotations.

Since the hybridization function is diagonal,
the negative sign problem arises from the competition between $t$ and $U$.
The system exhibits no sign problem in the two limiting cases $t=0$ and $U=0$. 
In these two cases, the impurity problem is diagonalized by choosing $\boldsymbol{U}$ to be the identity matrix or the matrix made of eigenvectors of $\mathcal{H}_\mathrm{imp}$, respectively.
We call the former basis the {\it site basis} and the latter basis the {\it diagonal basis}.

To find the optimal basis in terms of the average sign,
we perform a brute-force search in the parameter space of $\boldsymbol{U}$, restricting ourselves to (real) orthogonal matrices $\in SO(3)$ which can be parameterized by three Euler angles.
The average MC sign is computed on a uniform grid of $20\times 20\times 20$ for the Euler angles.
Throughout this section, we set $U=5$ and vary $t$ and $t^\prime$.

\subsubsection{Results for $t=t^\prime$}
\begin{figure}
    \begin{minipage}{\hsize}
	\centering\includegraphics[width=\textwidth,clip]{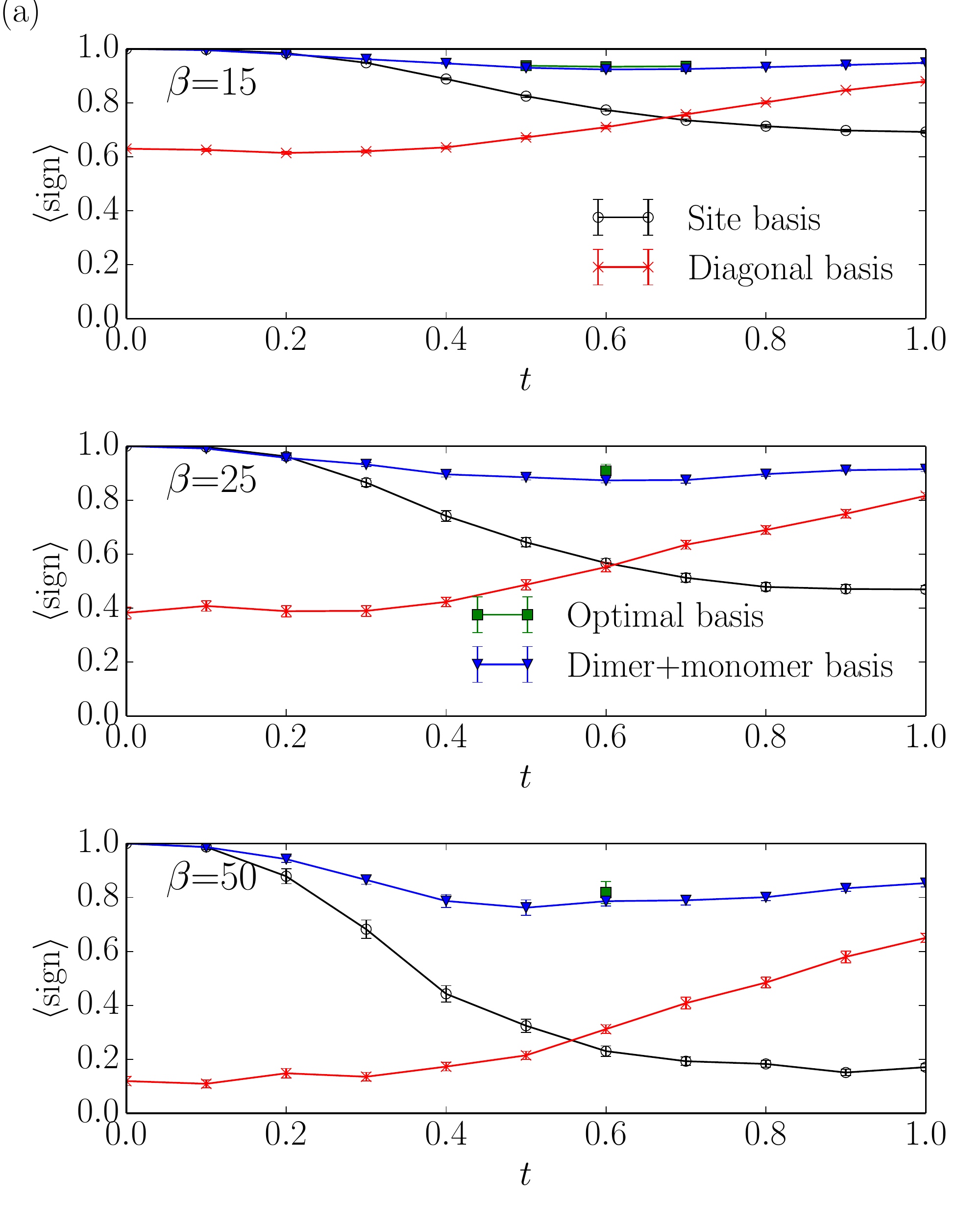}
	\centering\includegraphics[width=\textwidth,clip]{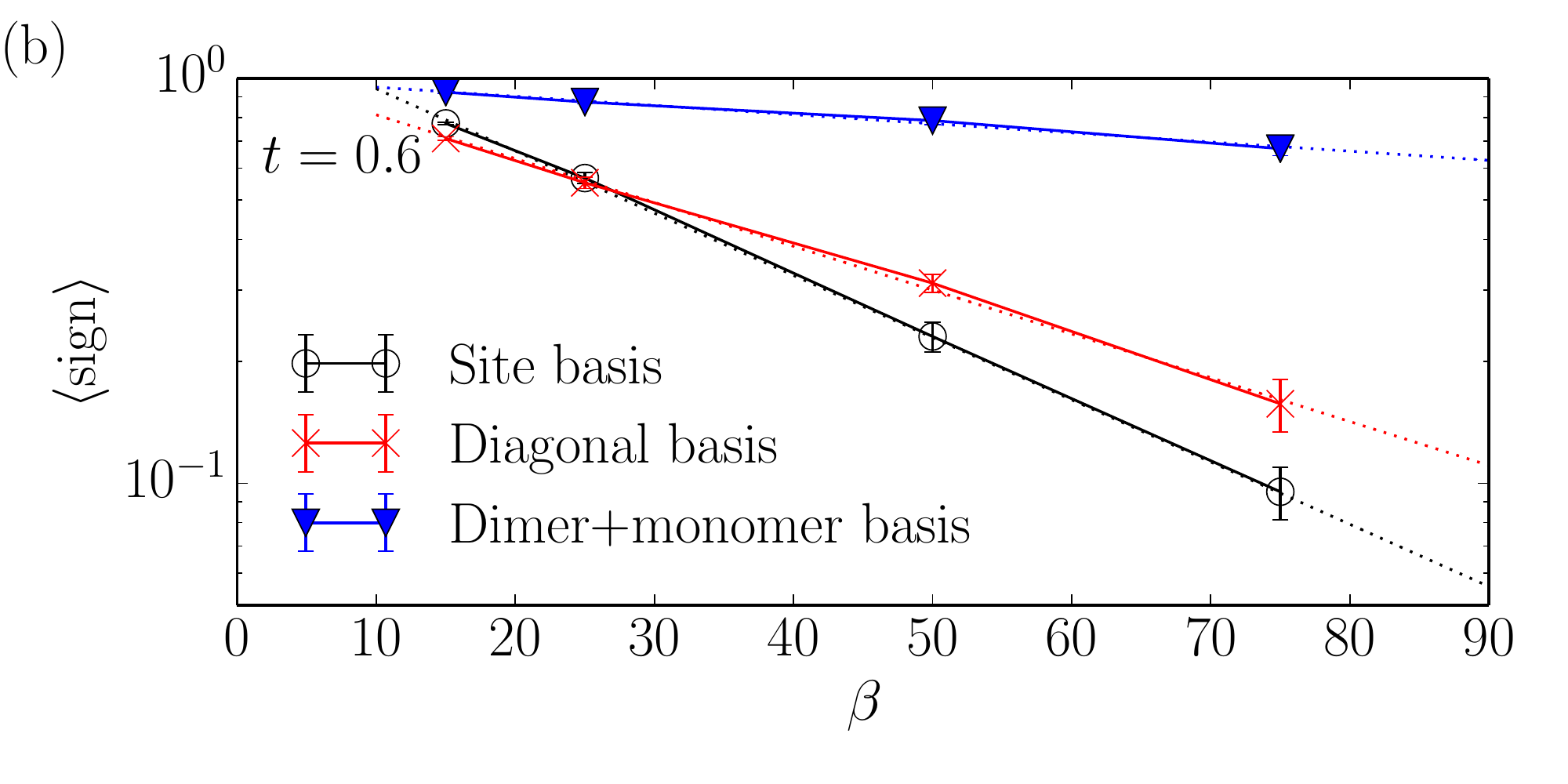}
	\caption{
		(Color online) 
		Average sign computed for the trimer model with $t^\prime/t=1$ and $U=5$ ($\beta=15,~25,~50$). 
		(a) $t$ dependence of the average sign.
		We show data for the site basis, the diagonal basis, the optimal basis found in the brute-force search, and the dimer+monomer basis.
		(b) $\beta$ dependence of the average sign computed for the site basis, the diagonal basis, and the dimer+monomer basis at $t=0.6$.
		The dotted lines denote a fit by the exponential form (\ref{eq:exp}).
	}
	\label{fig:trimer-sign}
	\end{minipage}
    \begin{minipage}{\hsize}
	\centering\includegraphics[width=\textwidth,clip]{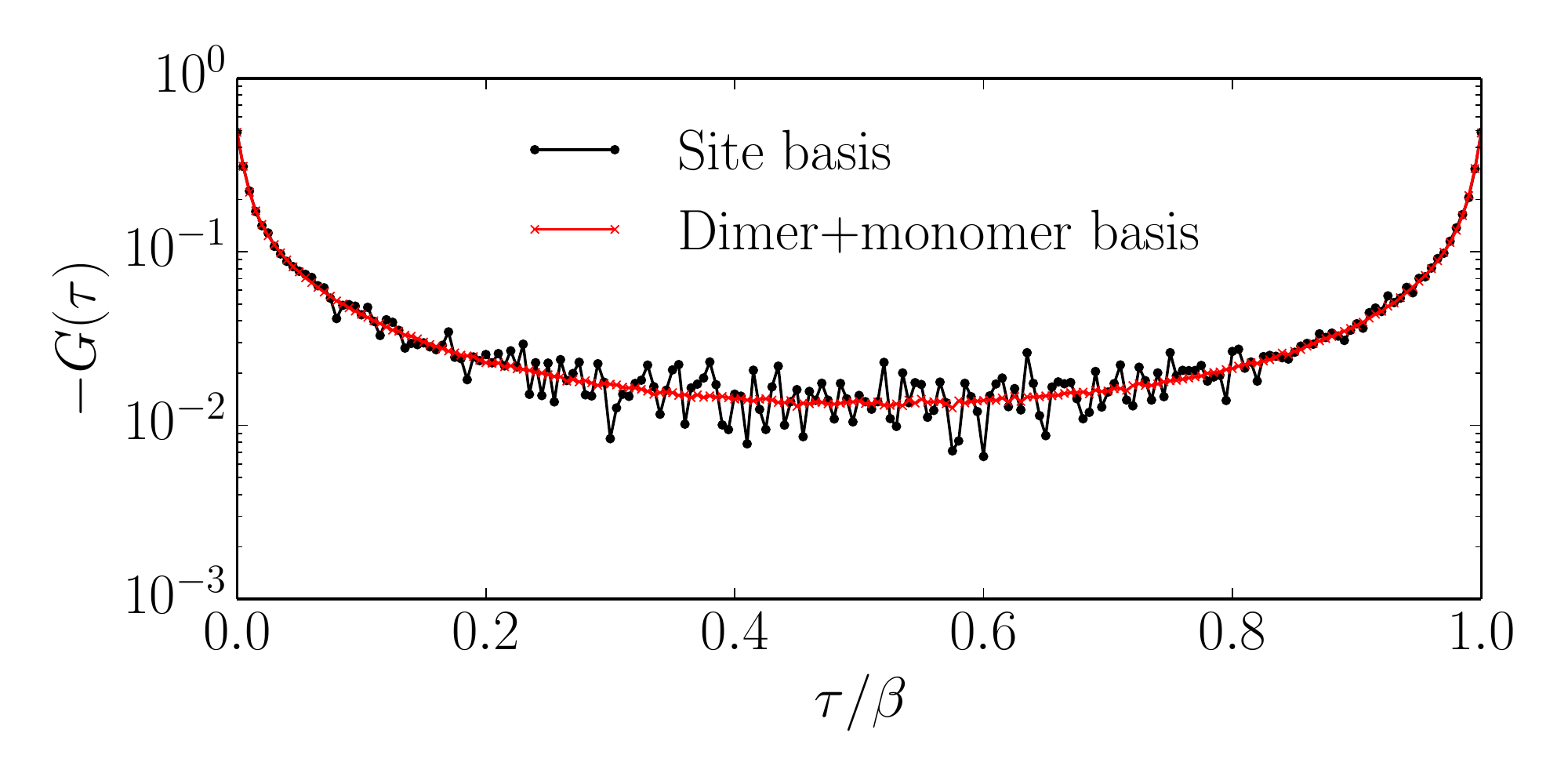}
	\caption{
		(Color online) 
		On-site imaginary-time Green's function $G(\tau)$ computed for the trimer model with $t=t^\prime=0.5$ and $U=5$ ($\beta=50$).
		We symmetrize the data using the three-fold rotational symmetry.
	}
	\label{fig:trimer-gf}
	\end{minipage}
\end{figure}

First, we discuss the results for $t=t^\prime$.
Figure~\ref{fig:trimer-sign} shows the average sign computed for the site basis and the diagonal basis as a function of $t$ for $U=5$ ($\beta=15,~25,~50$).
We also show the highest average sign found in the brute-force search for several parameters.
At $\beta=15$, the average signs for the site and diagonal bases cross at $t\simeq 0.7$.
Although the sign problem is severe for the site and diagonal bases at $t\simeq 0.7$,
the optimal basis found by the brute-force search has a considerably higher average sign with weaker $\beta$ dependence.
We found that the optimal basis corresponds to the transformation matrix 
\begin{eqnarray}
 U &=& 
 \left(
 \begin{array}{ccc}
 \frac{1}{\sqrt{2}} & \frac{1}{\sqrt{2}} & 0 \\ 
 \frac{1}{\sqrt{2}} & -\frac{1}{\sqrt{2}} & 0 \\ 
 0 & 0 & 1
 \end{array}
 \right),\label{eq:dimer+monomer}
 \end{eqnarray}
and therefore consists of bonding and anti-bonding orbitals on one of the three edges (1 and 2) and a localized orbital on the remaining site (3).
We call this basis the dimer+monomer basis.
(There are three equivalent optimal bases for $t=t^\prime$.)

\begin{figure}
	\centering\includegraphics[width=.3\textwidth,clip]{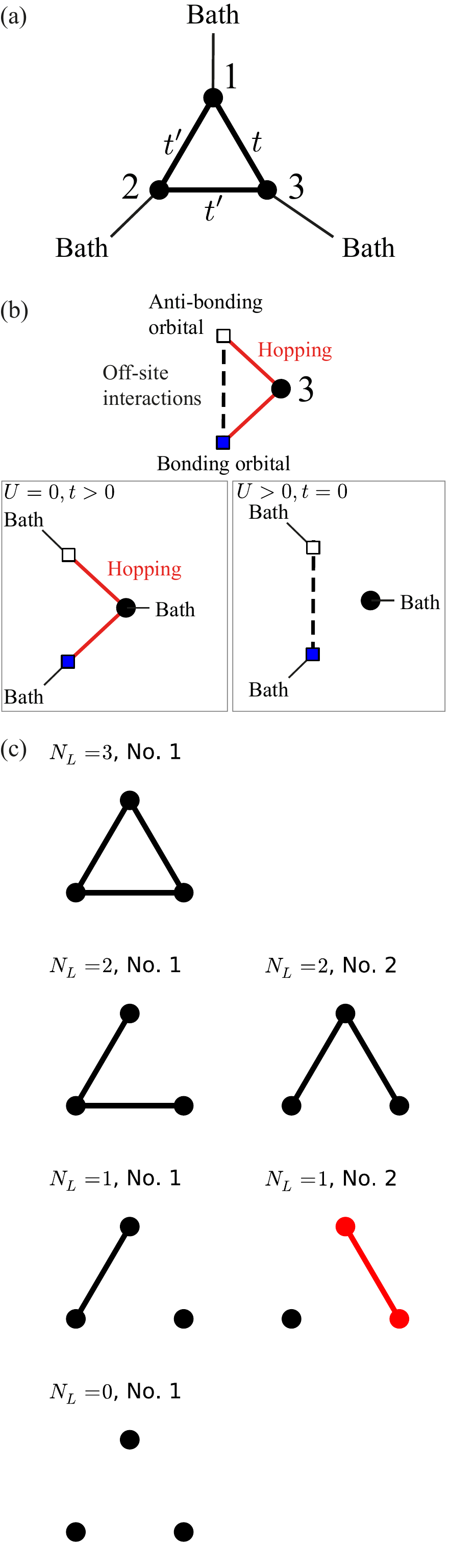}
	\caption{
		(Color online) 
		(a) Trimer impurity model with orbital-diagonal bath.
		Each site is connected to a bath with a semicircular density of states of width $4t$.
		(b) Schematic representation of the hoppings and the Coulomb interactions in the dimer+monomer basis ($t=t^\prime$).
		(c) Inequivalent realizations of links for the trimer impurity problem ($t\neq t^\prime$). $N_L$ is the number of links in a graph.
	}
	\label{fig:trimer}
\end{figure}

Figure~\ref{fig:trimer-sign} also shows the average sign for the dimer+monomer basis computed for $0 \le t \le 1$.
This basis has a higher average sign than the site and diagonal bases in this entire parameter region.
Another interesting observation is that the dimer+monomer basis is negative-sign-free in the two limiting cases: $t\gg U$ and $t=0$.
This may be because the system has no fermionic loop in the dimer+monomer basis for these two limiting cases.
As illustrated in Fig.~\ref{fig:trimer}(b),
there is no hopping  between the bonding and anti-bonding orbitals by construction.
Instead, the basis rotation generates off-site Coulomb interactions between them (including non-density-density terms).
The explicit form is given in Appendix~\ref{sec:coulomb-bonding}.
When $t=0$ and/or $U=0$,
the system contains no fermionic loop as is clearly seen in Fig.~\ref{fig:trimer}(b).
This gives an important insight, namely that the optimal 
basis diagonalizes some subset of the intracluster hoppings.
This allows to interpolate between the site and diagonal bases which are optimal in the limits $t=0$ and $U=0$.

Figure~\ref{fig:trimer-sign}(b) shows the $\beta$ dependence of the average sign computed for the site basis, the diagonal basis, and the dimer+monomer basis at $t=0.6$.
These data fit the exponential form
\begin{eqnarray}
  \braket{\mathrm{sign}} &\propto& \exp\left(-\frac{\beta}{\beta_\mathrm{sign}}\right),\label{eq:exp}
\end{eqnarray}
with $\beta_\mathrm{sign} = 28 \pm 1$, $40\pm 2$, $192\pm 13$, respectively.
For the dimer+monomer basis, the average sign  $\langle \mathrm{sign} \rangle$ decays about five times more slowly with respect to $\beta$ than for the diagonal basis.

In Fig.~\ref{fig:trimer-gf}, we compare the imaginary-time Green's function $G (\tau)$ computed for $t=0.5$ and $\beta=50$ using the site basis and the dimer+monomer basis
and the same number of MC steps. 
We plot the on-site (diagonal) element of $G(\tau)$ in the site basis.
Both results agree within statistical errors (indicated by the noise), but the errors of the dimer+monomer basis simulation are much smaller due to a larger average sign.

\begin{figure}
	\centering\includegraphics[width=.5\textwidth,clip]{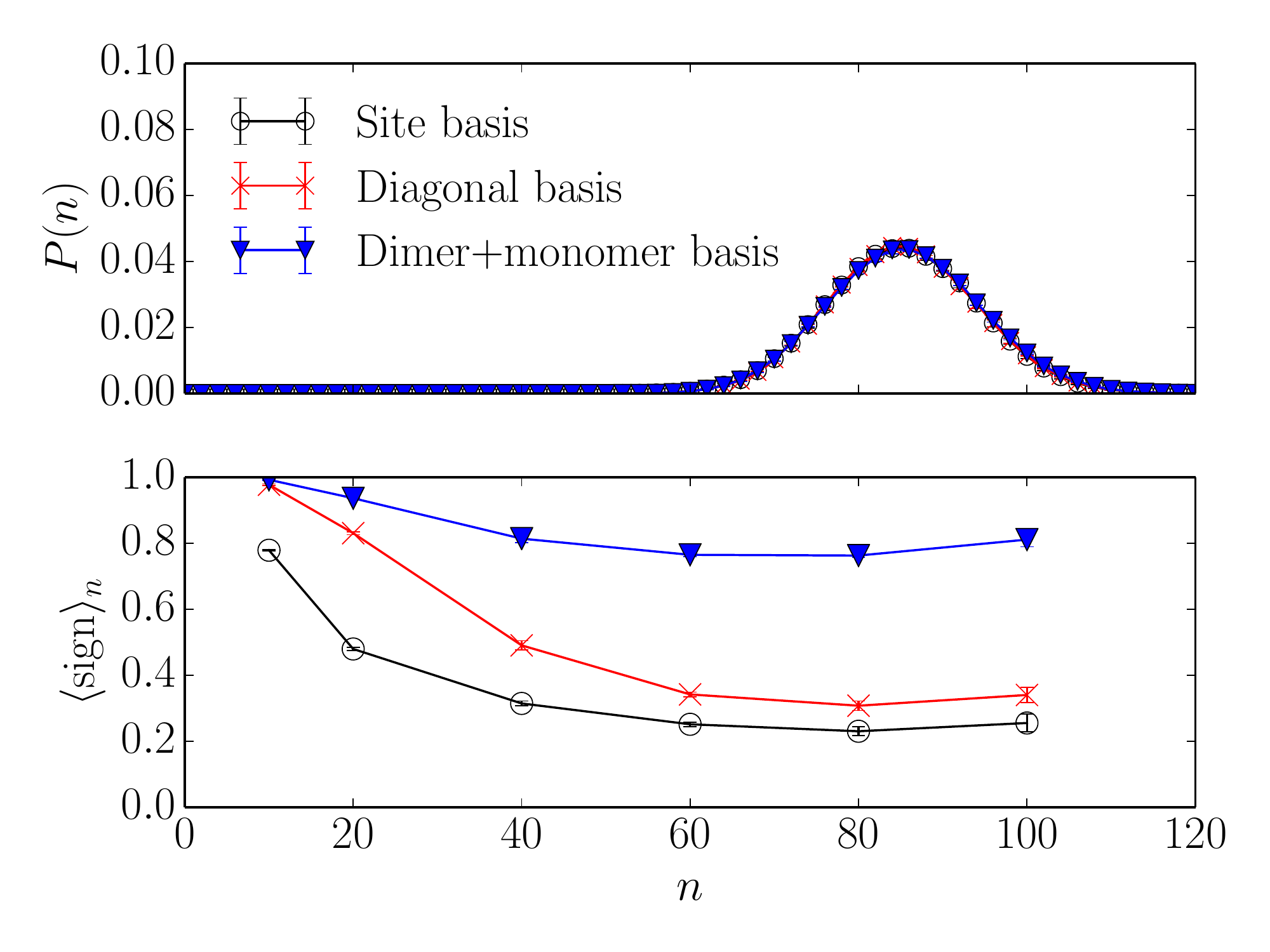}
	\caption{
		(Color online) 
		Distribution function of the expansion order ($n$) and expansion-order-resolved average sign computed at $t=t^\prime=0.6$ and $\beta=50$.
	}
	\label{fig:trimer-sign-order}
\end{figure}

Figure~\ref{fig:trimer-sign-order} shows the expansion-order-resolved average sign $\langle \mathrm{sign}\rangle_n$ computed at $t=t^\prime=0.6$ and $\beta=50$ using the site basis, the diagonal basis, and the dimer+monomer basis.
The distribution function of the expansion order has a single peak around $n=85$ independently of the choice of the single-particle basis.
On the other hand, the average sign $\langle \mathrm{sign}\rangle_n$ decreases monotonically for $n \le 85$ for these three bases, becoming almost independent of $n$ around the peak of the distribution function.
The superiority of the dimer+monomer basis is already discernible at low expansion orders $n\simeq 20$.
Note that the distribution function $P(n)$ generally depends on the single-particle basis used to represent the impurity.
In Fig.~\ref{fig:trimer-sign-order},
however, $P(n)$ seems to be almost identical for these three bases.
As detailed in Appendix~\ref{sec:P}, $P(n)$ is representation independent when $\langle \mathrm{sign}\rangle_n$ does not depend on $n$ even if $\langle \mathrm{sign}\rangle_n$ itself depends on the single-particle basis.
Our observation for $P(n)$ is consistent with the small $n$ dependence of $\langle \mathrm{sign} \rangle_n$ around $n\simeq 85$. 
\begin{figure*}
	\begin{tabular}{cc}
	\begin{minipage}{0.45\hsize}
	\centering\includegraphics[width=\textwidth,clip,type=pdf,ext=.pdf,read=.pdf]{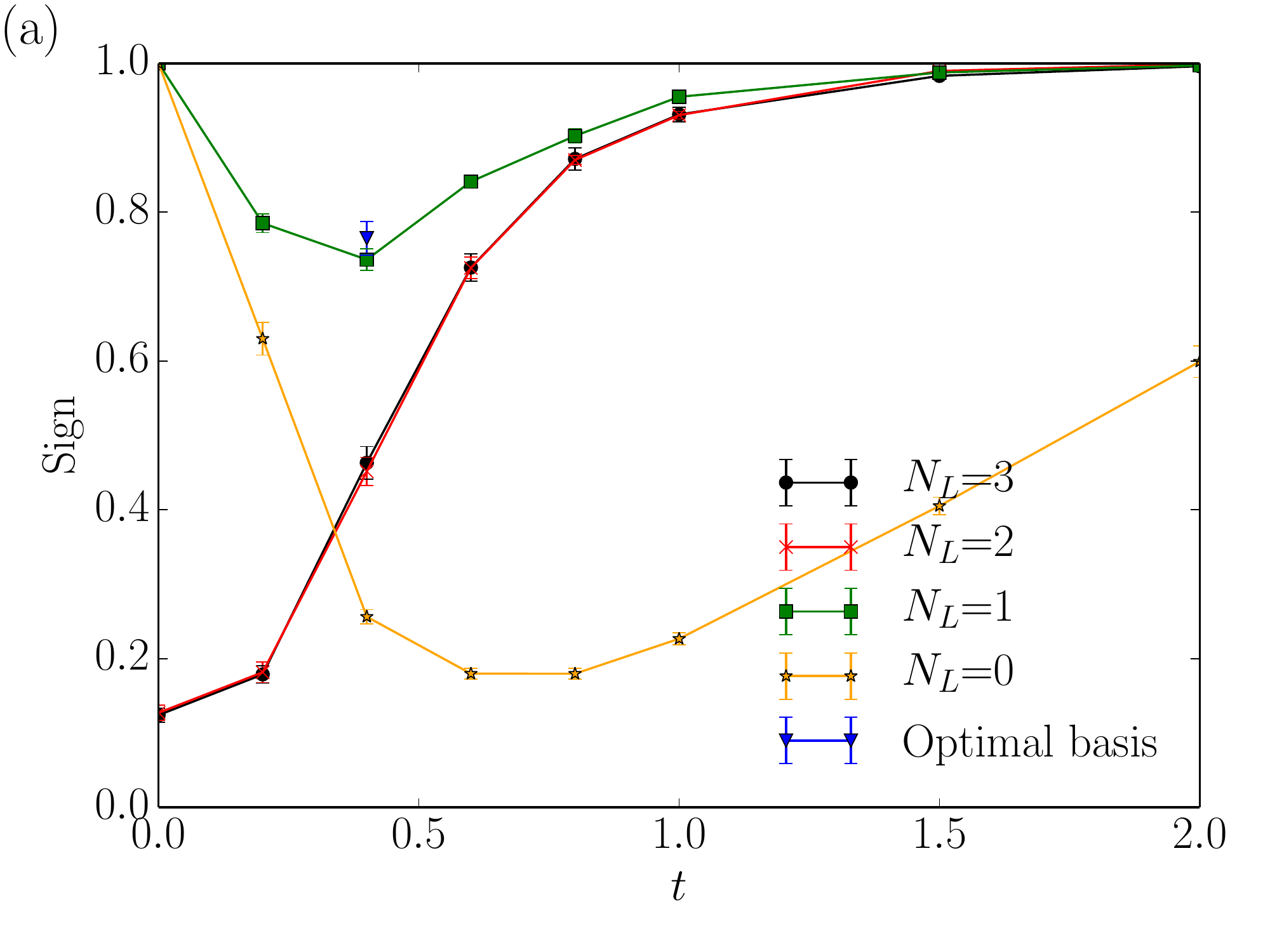}
	\end{minipage}
	&
	\begin{minipage}{0.45\hsize}
		\centering\includegraphics[width=\textwidth,clip,type=pdf,ext=.pdf,read=.pdf]{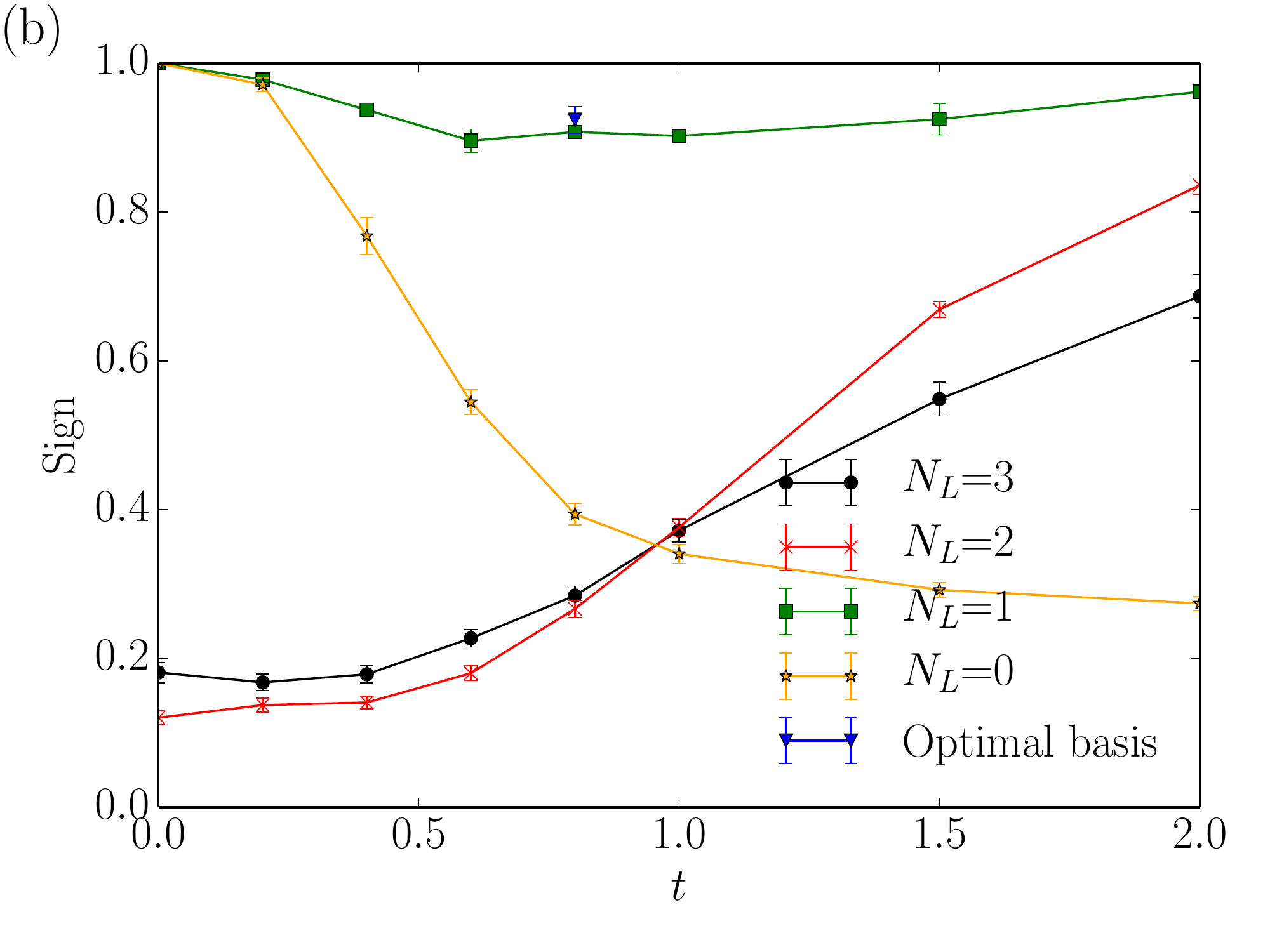}
	\end{minipage}	
    \end{tabular}
	\caption{
	(Color online) Average sign computed for the trimer model: (a) $t^\prime/t=2$ and (b) 0.5 at $\beta=50$.
	We show the maximum value of the average sign among single-particle bases generated by graphs with the same $N_L$.
	We also plot the highest value of the average sign found in a brute force search. 
	We show the average sign computed for each basis in Fig.~\ref{fig:trimer-sign-nosymm-all} of the Appendix.
	}
	\label{fig:trimer-sign-nosymm}
\end{figure*}

\subsubsection{Results for $t\neq t^\prime$}
We next discuss the results for $t \neq t^\prime$.
The result for $t=t^\prime$ suggested a guiding principle for generating good single-particle bases: diagonalizing a subset of the local hoppings.
We thus introduce the concept of \textit{partial diagonalization}, i.e. 
we define a new basis as the eigenbasis of some selected 
matrix elements of the intracluster single-particle Hamiltonian. 
%elements in the local hopping matrix.
In practice, the trimer impurity model has three edges (links) with %finite 
local hoppings.
We deactivate some of them for the purpose of partially diagonalizing the intracluster single-particle Hamiltonian. 
Figure~\ref{fig:trimer}(c) shows all symmetrically inequivalent graphs for $t\neq t^\prime$.
$N_L$ is the number of links remaining in a graph.
The graphs for $N_L=0$ and $3$ give the site basis and the diagonal basis, respectively.
Two dimer+monomer-type bases are generated from $N_L=1$ graphs.
The eigenbases are generated by diagonalizing the intracluster single-particle Hamiltonians of the graphs numerically.
Since some of the graphs have degenerate eigenvalues, we specify the unitary matrices used for the present study in Sec.~\ref{sec:matrix}.

We computed the average sign for these bases at $t^\prime/t=0.5$ and 2.0 ($\beta=50$).
Figures~\ref{fig:trimer-sign-nosymm}(a) and~\ref{fig:trimer-sign-nosymm}(b)
 show the highest average sign among the graphs with the same $N_L$ for $t^\prime/t=0.5$ and 2.0.
For the site and diagonal bases, the sign problem is severe around $t=0.4$ and 1.0, respectively.
In these regions, the highest average sign is produced by $N_L=1$ (No. 2), where a dimer is placed on the bond connecting 
sites 1 and 3.
This basis is always superior to $N_L=1$ (No. 1).

\subsection{Tetramer impurity model}
\subsubsection{Set-up}
\begin{figure*}
	\centering\includegraphics[width=\textwidth,clip]{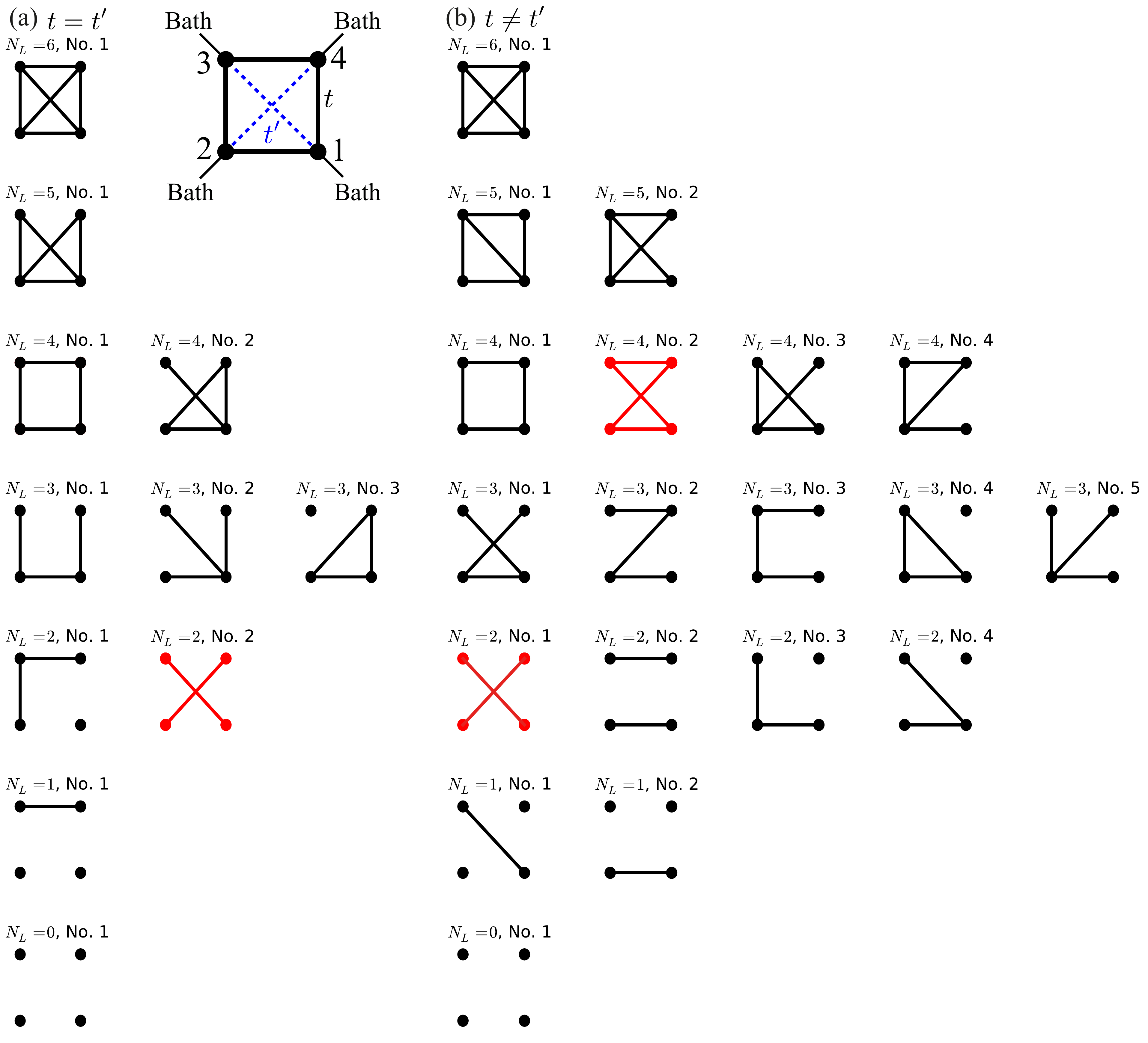}
	\caption{
		(Color online) 
		Tetramer impurity model (inset) and symmetrically inequivalent graphs for $t=t^\prime$ [(a)] and $t\neq t^\prime$ [(b)].
	}
	\label{fig:tetramer}
\end{figure*}

The tetramer impurity model is illustrated in Fig.~\ref{fig:tetramer}.
Its Hamiltonian reads
\begin{eqnarray}
 \mathcal{H} &=& -\sum_{i\neq j}^4 t_{ij} \opcdag_{i\sigma} \opc_{j\sigma} + U\sum_{i=1}^4 \hat{n}_{i\uparrow} \hat{n}_{i\downarrow}-\mu\sum_{i=1}^4 \hat{n}_i\nonumber\\
  && + \lambda \sum_{i=\alpha}\sum_{k\sigma} (\opcdag_{i\sigma} a_{\alpha k \sigma} + a^\dagger_{\alpha k \sigma} \opc_{i\sigma} )\nonumber\\
  && + \sum_{\alpha} \sum_{k \sigma} \epsilon_k a^\dagger_{\alpha k \sigma} a_{\alpha k \sigma},
\end{eqnarray}
where $\mu=U/2$ and $\lambda=1$.
Throughout this section, we measure energy in units of $\lambda$.
As in the case of the trimer impurity model,
each site is connected to a bath with a semicircular density of states of width 4.
That is, the distribution of their energy levels $\epsilon_k$ is given by $\sum_k \delta (\epsilon - \epsilon_k) = \frac{1}{2 \pi} \sqrt{4 - \epsilon^2}$.
The definition of the matrix elements $\{t_{ij}\}$ is shown in Fig.~\ref{fig:tetramer}(a).
When $t=t^\prime$, this model is equivalent to a tetrahedron impurity model with cubic symmetry.
The ratio $t^\prime/t~(\le 1)$ controls the strength of the geometrical frustration.
Throughout this section, we set $U=5$ and vary $t$ and $t^\prime$.
\begin{figure*}
	\begin{tabular}{cc}
		\begin{minipage}{0.45\hsize}
			\centering\includegraphics[width=\textwidth,clip,type=pdf,ext=.pdf,read=.pdf]{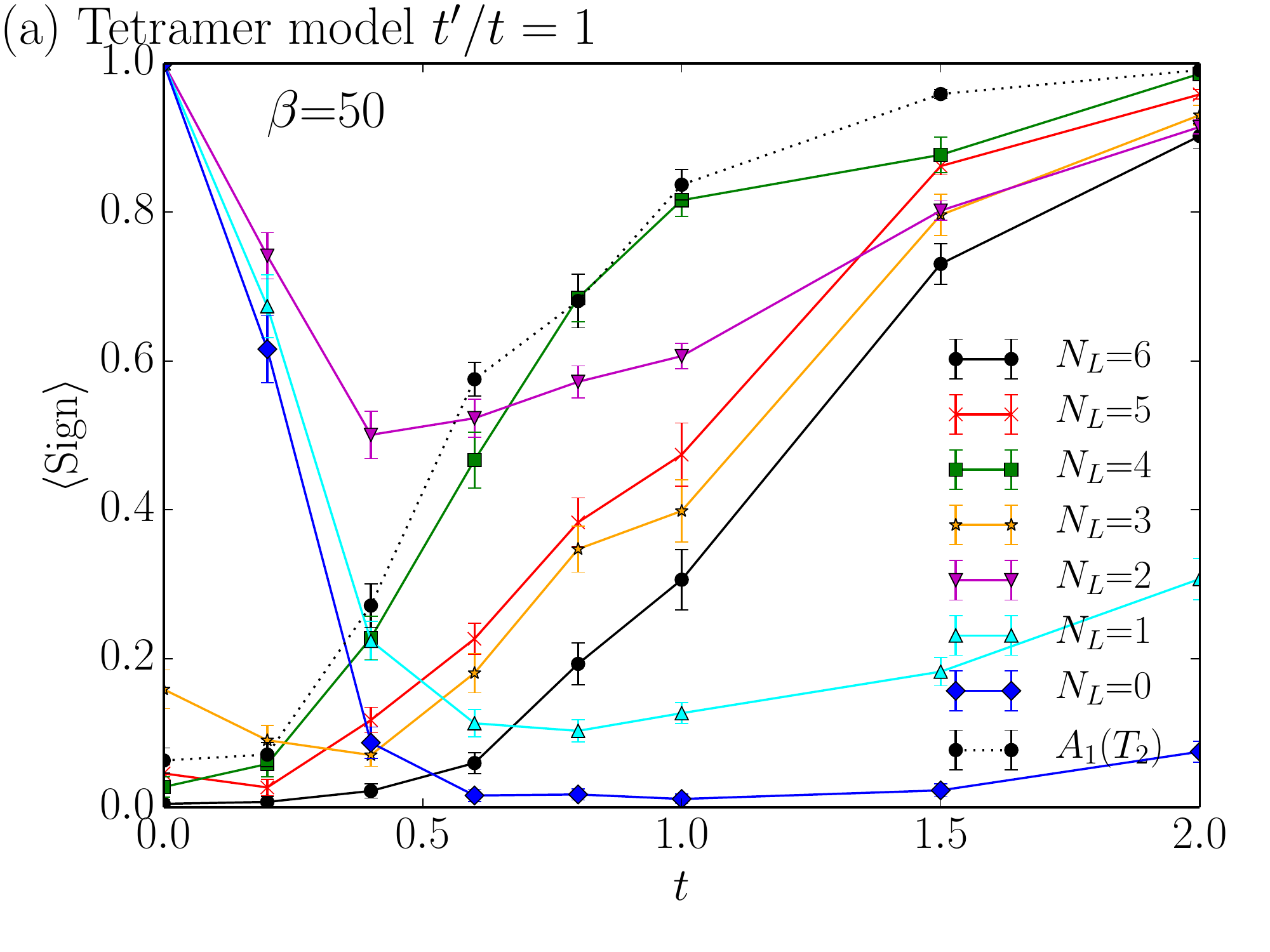}
			\centering\includegraphics[width=\textwidth,clip,type=pdf,ext=.pdf,read=.pdf]{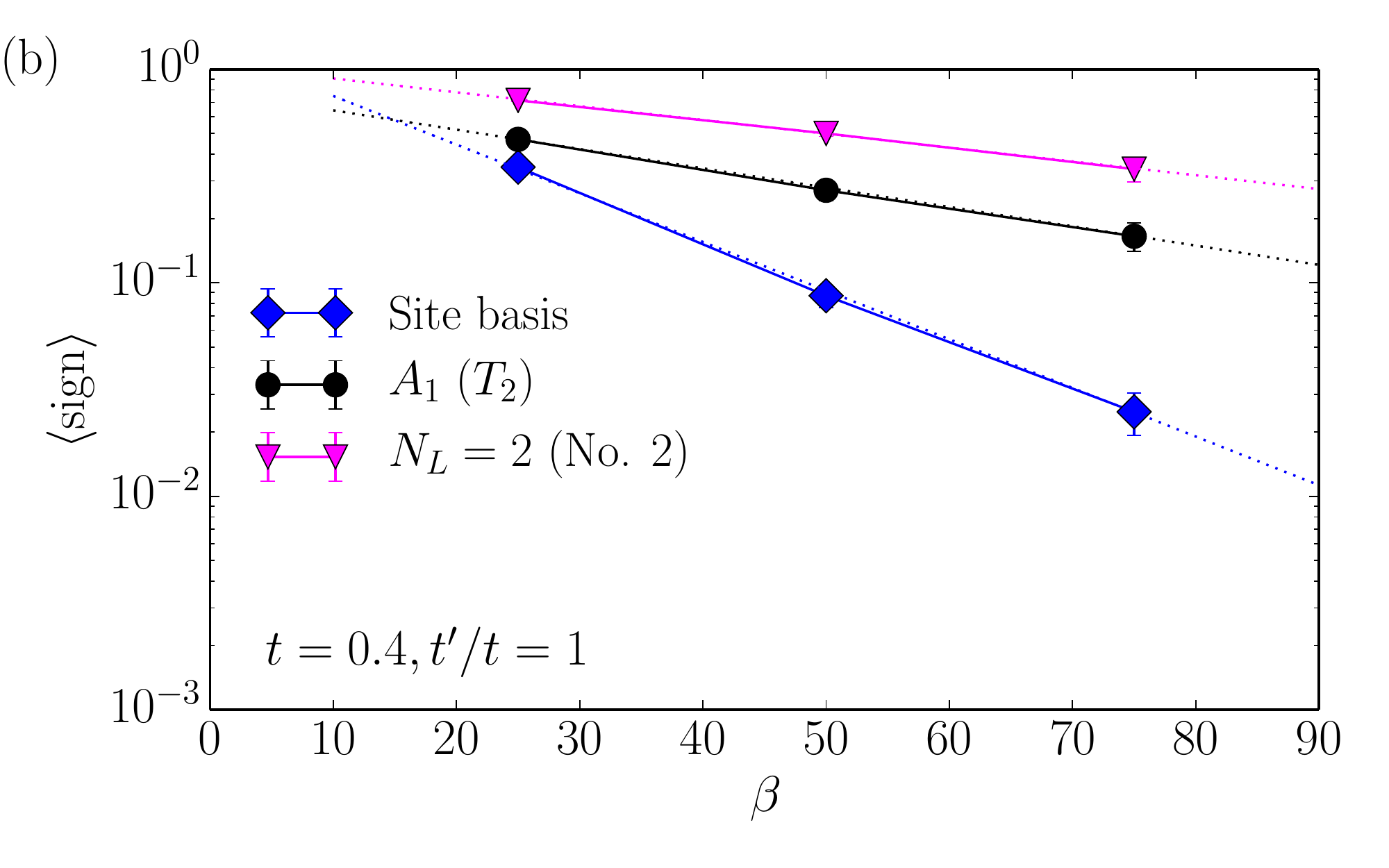}
		\end{minipage}
		&
		\begin{minipage}{0.45\hsize}
			\centering\includegraphics[width=\textwidth,clip,type=pdf,ext=.pdf,read=.pdf]{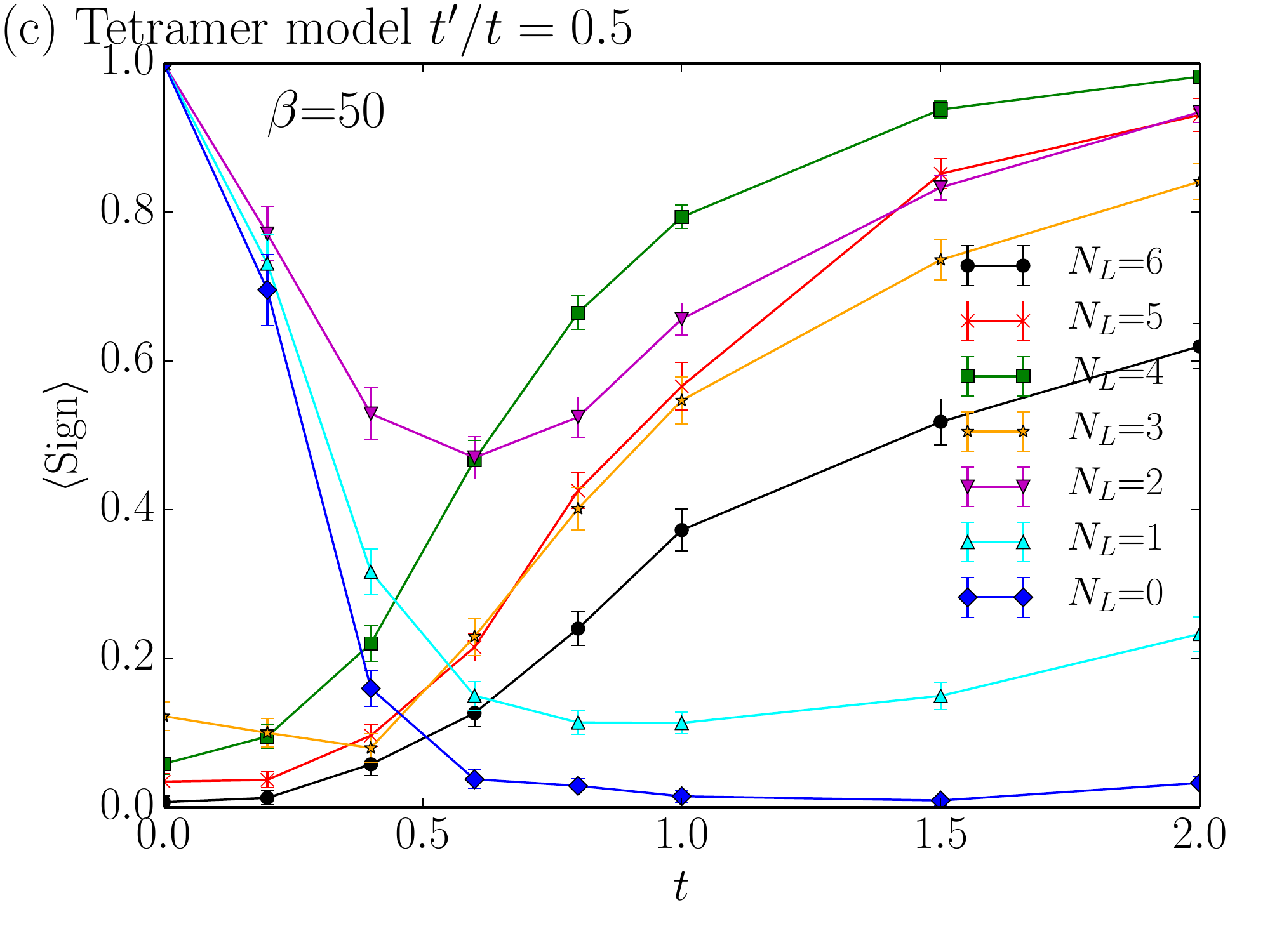}
			\centering\includegraphics[width=\textwidth,clip,type=pdf,ext=.pdf,read=.pdf]{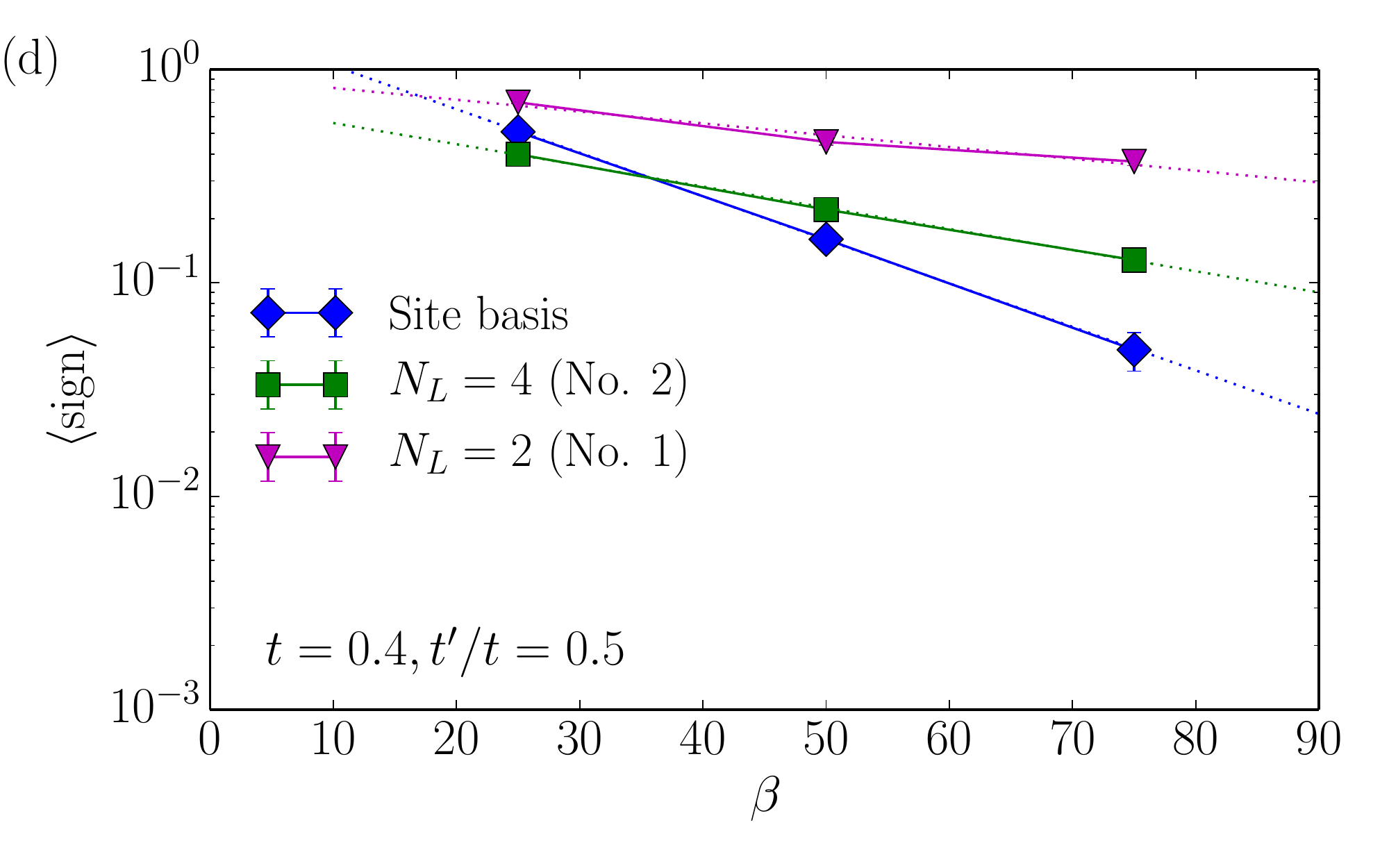}
		\end{minipage}	
	\end{tabular}
	\caption{
		(Color online) Average sign computed for the tetramer model:
		(a)/(b) $t^\prime/t=1$ and (c)/(d) $t^\prime/t=0.5$.
		In (a) and (c), we show the maximum value of the average sign among single-particle bases generated by graphs with the same $N_L$ computed at $\beta=50$.
		In (b) and (d), we show the $\beta$ dependence of the average sign computed at $t=0.4$.
        The dotted lines are fits by the exponential~(\ref{eq:exp}).
	}
	\label{fig:tetramer-sign}
\end{figure*}

\subsubsection{Results for $t=t^\prime$}
Let us start by considering the results for $t=t^\prime$.
Figure~\ref{fig:tetramer}(a) shows all symmetrically inequivalent graphs for $t=t^\prime$.
We obtained unitary matrices by numerically diagonalizing the corresponding intracluster single-particle Hamiltonians.~\footnote{We used the numpy package for Python.}
Some graphs have degenerate eigenvalues and hence the unitary matrix is not uniquely determined.
For example, we obtained
\begin{eqnarray}
	U &=& 
	\left(
	\begin{array}{cccc}
-1/2  &  -\frac{3}{2 \sqrt{3}}  &  0  &  0 \\
-1/2  &  \frac{1}{2 \sqrt{3}}  &  -\frac{1}{\sqrt{3}}  &  -\frac{1}{\sqrt{3}} \\ 
-1/2  &  \frac{1}{2 \sqrt{3}}  &  \frac{1}{2} + \frac{1}{2 \sqrt{3}}  &  - \frac{1}{2} + \frac{1}{2 \sqrt{3}} \\ 
-1/2  &  \frac{1}{2 \sqrt{3}}  &  - \frac{1}{2} + \frac{1}{2 \sqrt{3}}  &  \frac{1}{2} + \frac{1}{2 \sqrt{3}}  \\
	\end{array}
	\right)
\end{eqnarray}
with the eigenvalues $(-3, 1, 1, 1)$ for $N_L=6$ (No. 1).
We provide all these unitary matrices in Appendix~\ref{sec:matrix}.
In addition, we test the $A_1$ ($T_2$) representations of the $T_d$ point group,~\cite{Hattori:2012bh}
\begin{eqnarray}
	U &=& \frac{1}{2}
	\left(
	\begin{array}{cccc}
		 1    &     1 &   1  &  1       \\
		 1    &    -1 &  -1  &  1       \\
		 1    &     1 &  -1  & -1       \\
		 1    &    -1 &   1  & -1       \\
	\end{array}
	\right),\label{eq:A1T2}
\end{eqnarray}
which diagonalizes the intracluster single-particle Hamiltonian of the tetramer as $N_L=6$ (No. 1) does.
We name this basis $A_1$ ($T_2$). 
$N_L=0$ corresponds to the site basis with $U$ being the identity matrix.

Figure~\ref{fig:tetramer-sign}(a) presents the average sign computed at $\beta=50$.
We show only the maximum values of the average sign among single-particle bases generated by graphs with the same $N_L$.
Refer to Fig.~\ref{fig:tetramer-sign-each-graph} in the Appendix for the data for each graph.
The system exhibits a severe sign problem for the site basis [$N_L=0$ (No. 1)], $N_L=6$ (No. 1), and $A_1$ ($T_2$) at $t\simeq 0.4$.
A notable point is that $A_1$ ($T_2$) is superior to 
%the diagonal basis
$N_L=6$ (No. 1) 
in the entire range of $t$.
The only difference between these two bases is in the structure of the Coulomb matrix.
At $t\simeq 0.4$, the highest average sign is given by one of our nontrivial bases, $N_L=2$ (No. 2) which consists of two independent dimers.
Its unitary matrix reads
\begin{eqnarray}
	U &=& 
	\left(
	\begin{array}{cccc}
		-\frac{1}{\sqrt{2}} & 0          & 0          &-\frac{1}{\sqrt{2}}\\
		0          &-\frac{1}{\sqrt{2}} & -\frac{1}{\sqrt{2}} & 0       \\
		-\frac{1}{\sqrt{2}} & 0          & 0           & \frac{1}{\sqrt{2}}\\
		0          &-\frac{1}{\sqrt{2}} & \frac{1}{\sqrt{2}}  & 0        \\
	\end{array}
	\right),\label{eq:two-dimers}
\end{eqnarray}
and the eigenvalues of the intracluster single-particle Hamiltonian of the graph are $(-1, -1, 1, 1)$, respectively.

On the other hand, the $N_L=4$ (No. 1) basis has a rather high average sign comparable to $A_1$ ($T_2$) overall.
While this basis is numerically constructed to diagonalize the intracluster single-particle Hamiltonian corresponding to $N_L=4$ (No. 1),
it is found to also diagonalize that of $N_L=6$, 
consisting of a different linear combination of the three-fold eigenvectors from 
%the diagonal basis 
$N_L=6$ (No. 1) 
and $A_1$ ($T_2$).
Although one may be able to perform a brute-force search for the best linear combination,
such a study is left for the future because of its high computational cost.

Figure~\ref{fig:tetramer-sign}(b) shows the $\beta$ dependence of the average sign computed for the site basis, $A_1$ ($T_2$), $N_L=2$ (No. 2) at $t=0.4$.
The data fit the exponential form~(\ref{eq:exp}) with $\beta_\mathrm{sign}=18.8 \pm 1$, $48\pm 4$, $67 \pm 6$, respectively.
The average sign decays most slowly for $N_L=2$ (No. 2) among these three.

\subsubsection{Results for $t<t^\prime$}
For $t<t^\prime$, there are more symmetrically inequivalent graphs, as shown in Fig.~\ref{fig:tetramer}(b), due to the absence of the cubic symmetry.
The $N_L=6$ (No. 1) basis 
diagonalizes the local hoppings of the $N_L=6$ graph with the eigenvalues of $(-2.5,0.5, 0.5, 1.5)$.
Some bases obtained for other graphs consist of different linear combinations of these doubly degenerate eigenvectors.
For example, for the loop-type graph $N_L=4$ (No. 2), we obtained the same transformation matrix as that of $A_1$ ($T_2$) [Eq.~(\ref{eq:A1T2})].
This transformation matrix also diagonalizes the intracluster single-particle Hamiltonian of $N_L=6$.

We illustrate the $t$ dependence of the average sign computed for $t^\prime/t=0.5$ in Fig.~\ref{fig:tetramer-sign}(c).
We show only the maximum values among graphs with the same $N_L$.
Refer to Fig.~\ref{fig:tetramer-sign-each-graph} in the Appendix for the data for each graph.
The trend is similar to $t^\prime/t=1$.
For larger $t$, the highest average sign is given by the loop-type graph $N_L=4$ (No. 2),
\footnote{We confirmed that $A_1$ ($T_2$) indeed has the highest average sign for $0.2\le t\le 1$ and $\beta=50$ among transformation matrices which diagonalize the intracluster single-particle Hamiltonian of $N_L=6$.}
while the two-dimer-type graph $N_L=2$ (No. 1) is optimal for smaller $t$.
We plot the $\beta$ dependence of the average sign at $t=0.4$ in Fig.~\ref{fig:tetramer-sign}(d).
The data fit the exponential form (\ref{eq:exp}) with
$\beta_\mathrm{sign}=21\pm 1$, $44\pm 2$, $78\pm 3$ for the site basis, $N_L=4$ (No. 2), and $N_L=2$ (No. 1), respectively.

\section{Interaction-expansion}

\subsection{Implementation}

Motivated by the results for CT-HYB, we study the single-particle basis dependence of the sign problem for interaction-expansion continuous-time Monte Carlo (CT-INT) where we expand the partition function in powers of the interaction terms. 
We refer to Ref.~\onlinecite{Gull:2011jda} for a review of CT-INT.
Even if the original model contains only on-site repulsion,
a single-particle basis rotation may generate non-density-density off-site interactions.
The model presented in the new basis can be regarded as a multi-orbital impurity problem.
However, for CT-INT, we have to expand in these interaction explicitly in contrast to CT-HYB.
Therefore, in this paper, we restrict ourselves to single-particle bases for which non-local interactions are of ``Slater-Kanamori'' type, i.e., inter-orbital repulsion, intra-orbital repulsion, exchange coupling, Hund coupling, and pair hopping.
This can be done by restricting ourselves to graphs consisting of independent dimer(s) such as the dimer+monomer basis for the trimer and $N_L=2$ (No. 2) for the tetramer shown in Fig.~\ref{fig:tetramer}(a).
For these bases, the Slater-Kanamori interaction with an unusual parameterization acts between each pair of a bonding orbital and an anti-bonding orbital as
\begin{equation}
\frac{1}{2}\sum_{\alpha\beta\alpha^\prime\beta^\prime}\sum_{\sigma\sigma^\prime}U_{\alpha\beta\alpha^\prime\beta^\prime}d^\dagger_{\alpha\sigma}d^\dagger_{\beta\sigma^\prime}d_{\beta^\prime\sigma^\prime}d_{\alpha^\prime\sigma},
\end{equation}
with the intra-orbital repulsion $U_{\alpha\alpha\alpha\alpha}=U/2$, the inter-orbital repulsion $U_{\alpha\beta\alpha\beta}=U^\prime=U/2$, the exchange interaction $U_{\alpha\beta\beta\alpha}=J_\mathrm{H}=U/2$, and the pair hopping $U_{\alpha\alpha\beta\beta}=J_\mathrm{H}^\prime=U/2$.
$\alpha$ and $\beta$ are the bonding/anti-bonding orbitals  ($\alpha\neq \beta$).
$\sigma$ and $\sigma^\prime$ are spin indices.
The derivation is given in Appendix~\ref{sec:coulomb-bonding}.
We treat these models by an efficient algorithm recently proposed by some of the authors~\cite{Nomura:2014by,Nomura:2015cb} with the further optimization described in Appendix~\ref{sec:double-vertex-update}.
While this transformation increases the number of interaction types, it does not increase the average perturbation order since the strength of each interaction decreases.
The Green function as a function of imaginary time is represented using 100 Legendre orthogonal polynomials.~\cite{Boehnke:2011dd}

\subsection{Trimer}
\begin{figure}
	\includegraphics[width=.49\textwidth,clip]{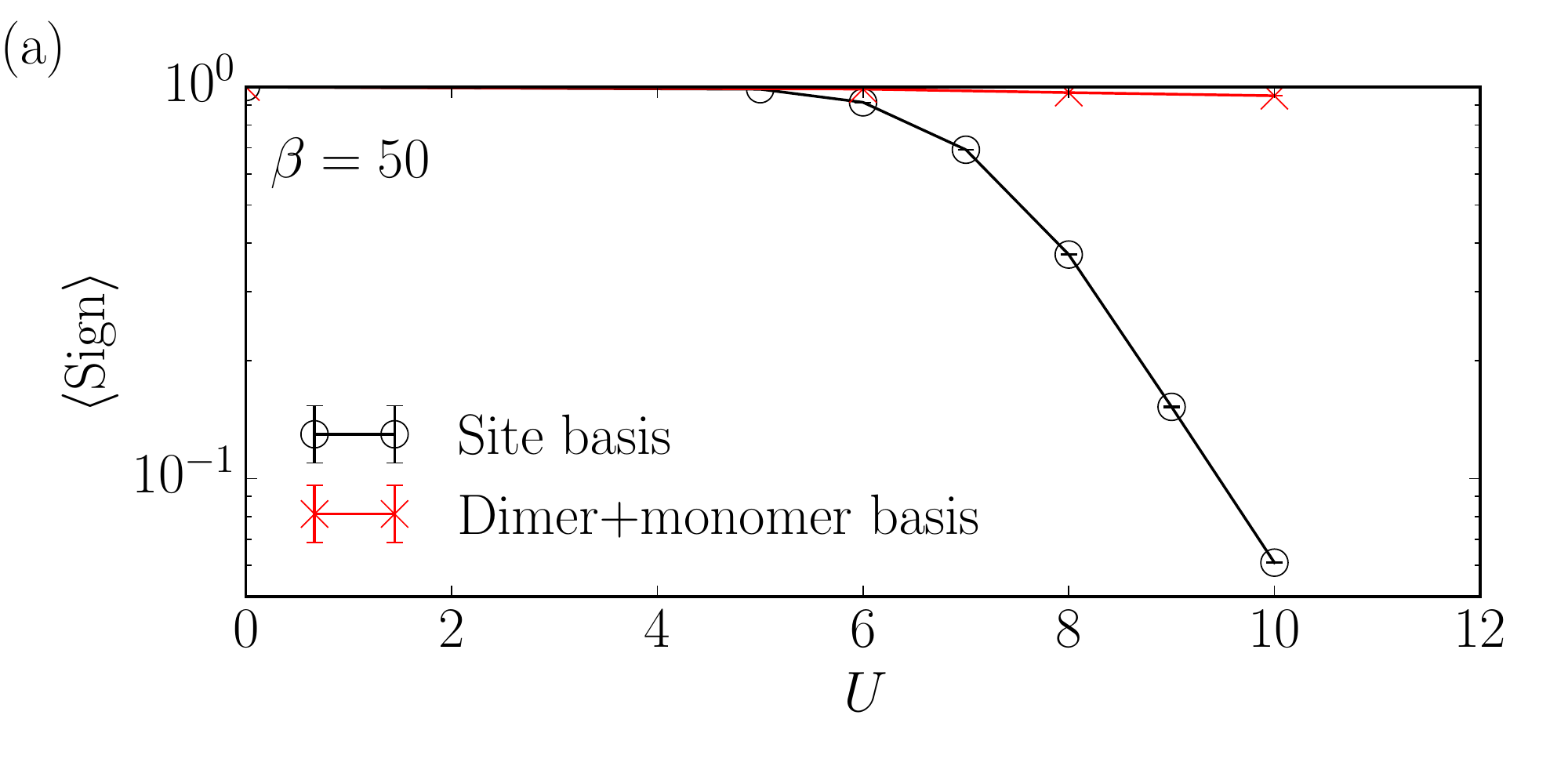}
	\includegraphics[width=.49\textwidth,clip]{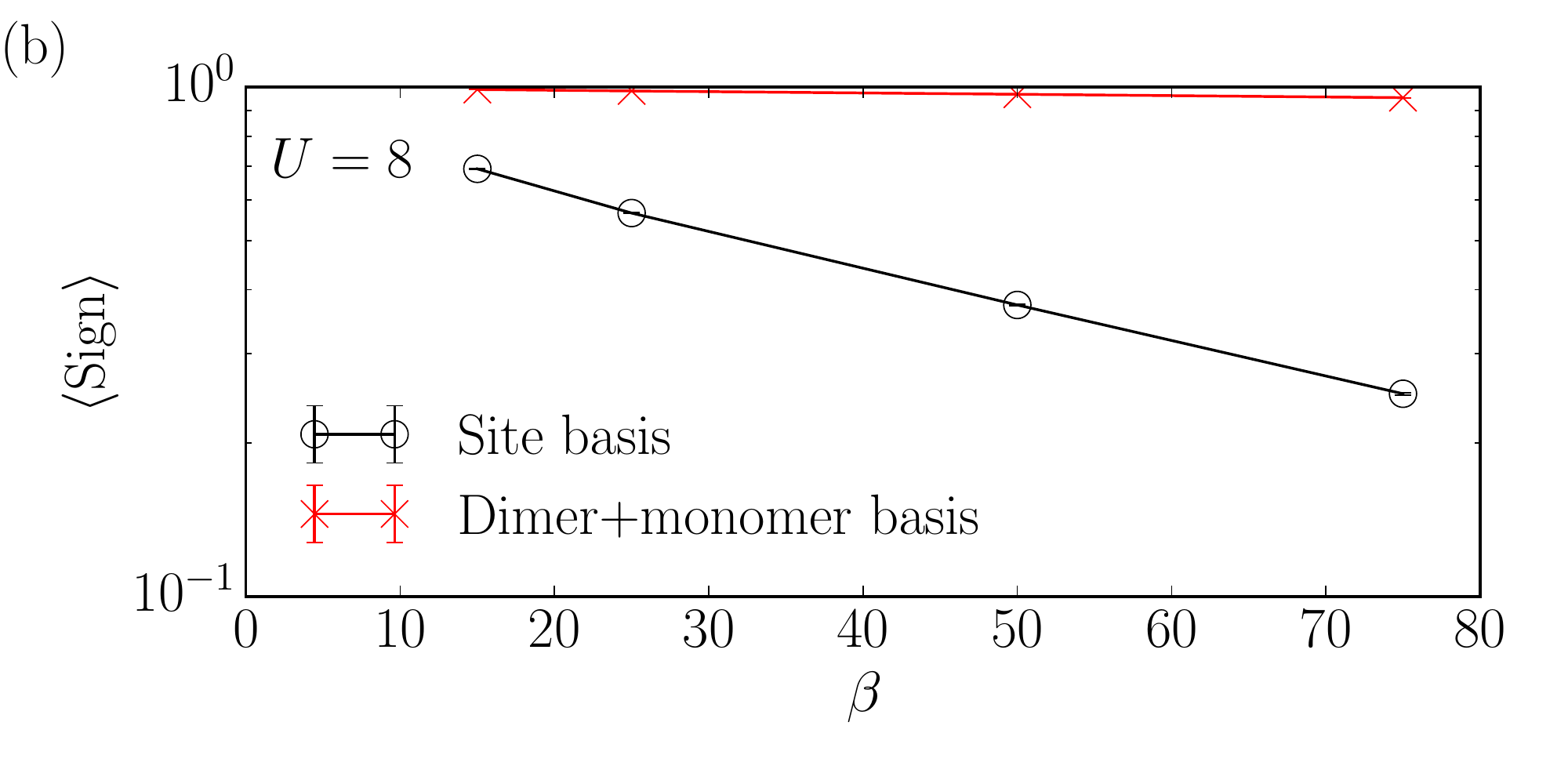}
	\caption{
		(Color online) 
		Comparison of the average sign computed for the trimer model using the site basis and the dimer+monomer basis by CT-INT.
		(a) $U$ dependence of the average sign at $\beta=50$.
		(b) $\beta$ dependence of the average sign at $U=8$.
	}
	\label{fig:trimer-ctint-sign}
\end{figure}

We first study the trimer impurity model (\ref{eq:trimer}) for $t=t^\prime=1$.
In Fig.~\ref{fig:trimer-ctint-sign}(a),
we show the $U$ dependence of the average sign computed using the site basis and the dimer+monomer basis at $\beta=50$.
For the site basis, we face a severe sign problem for $U\gtrsim 7$. 
On the other hand, the sign problem is substantially reduced for the dimer+monomer basis.
The average sign is about 0.95 even at $U=10$ for this basis.

Figure~\ref{fig:trimer-ctint-sign}(b) compares the $\beta$ dependence of the average sign for these two bases at $U=8$.
The average sign vanishes exponentially with increasing $\beta$.
Fitting the data by Eq.~(\ref{eq:exp}), we obtained $\beta_\mathrm{sign}=59.5\pm 0.3$ and $1634\pm 17$ for the site basis and the dimer+monomer basis, respectively.
This indicates that the dimer+monomer basis allows to reach about 28 times lower temperatures compared to the site basis.

\subsection{Tetramer}
\begin{figure}
	\includegraphics[width=.49\textwidth,clip]{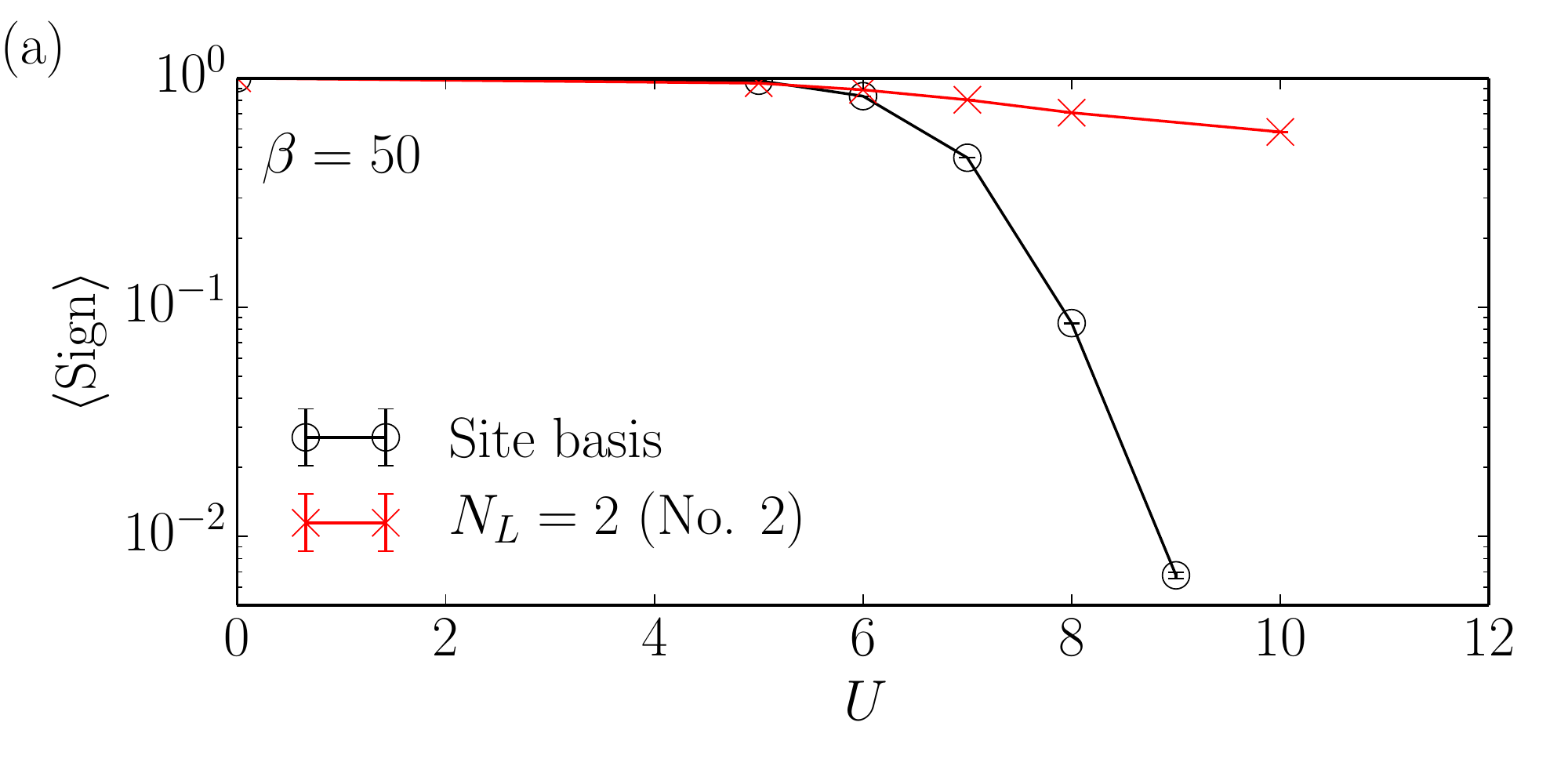}
	\includegraphics[width=.49\textwidth,clip]{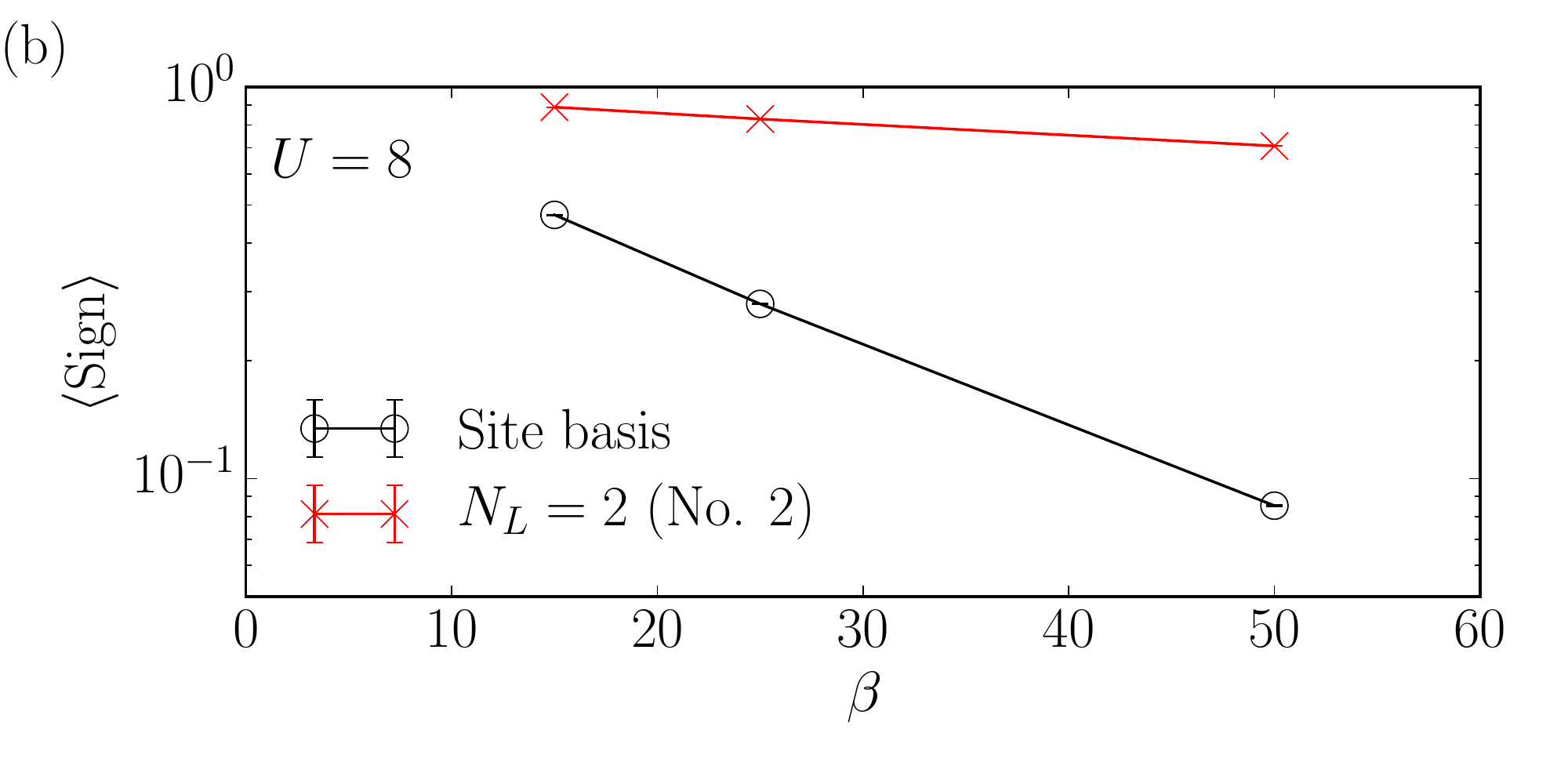}
	\caption{
		(Color online)
		Comparison of the average sign computed for the tetramer using the site basis and the $N_L=2$ (No. 2) basis by CT-INT.
		(a) $U$ dependence of the average sign at $\beta=50$.
		(b) $\beta$ dependence of the average sign at $U=8$.
	}
	\label{fig:tetramer-ctint-sign}
\end{figure}

For the tetramer ($t=t^\prime=1$),
we test $N_L=2$ (No. 2) given in Eq.~(\ref{eq:two-dimers}),
which consists of two independent pairs of bonding and anti-bonding orbitals.
In Fig.~\ref{fig:tetramer-ctint-sign}(a), we compare the $U$ dependence of the average sign computed for the site basis and $N_L=2$ (No. 2) at $\beta=50$.
For the site basis, one can see a severe sign problem for $U\gtrsim 6$.
Note that the band width of the localized tetramer, which is not connected to the bath, is 6.
On the other hand, for $N_L=2$ (No. 2), the average sign decreases substantially more slowly.
We plot the $\beta$ dependence of the average sign at $U=8$ in Fig.~\ref{fig:tetramer-ctint-sign}(b).
The average sign vanishes following the exponential form (\ref{eq:exp}) with $\beta_\mathrm{sign}=20.58\pm 0.07$ and $154.2\pm 0.9$ for the site basis and $N_L=2$ (No. 2), respectively.

To demonstrate the advantage of $N_L=2$ (No. 2), we compare the self-energy $\Sigma(\mathrm{i}\omega_n)$ computed by using the site basis and the $N_L=2$ (No. 2) basis at $U=8$ and $\beta=50$.
We used the same MC time for the site basis and $N_L=2$ (No. 2) to make a fair comparison.
Refer to Appendix~\ref{sec:double-vertex-update} for technical details.
The data obtained by the $N_L=2$ (No. 2) are substantially less noisy for both of the site-diagonal and site-off-diagonal elements.
\begin{figure}
	\includegraphics[width=.49\textwidth,clip]{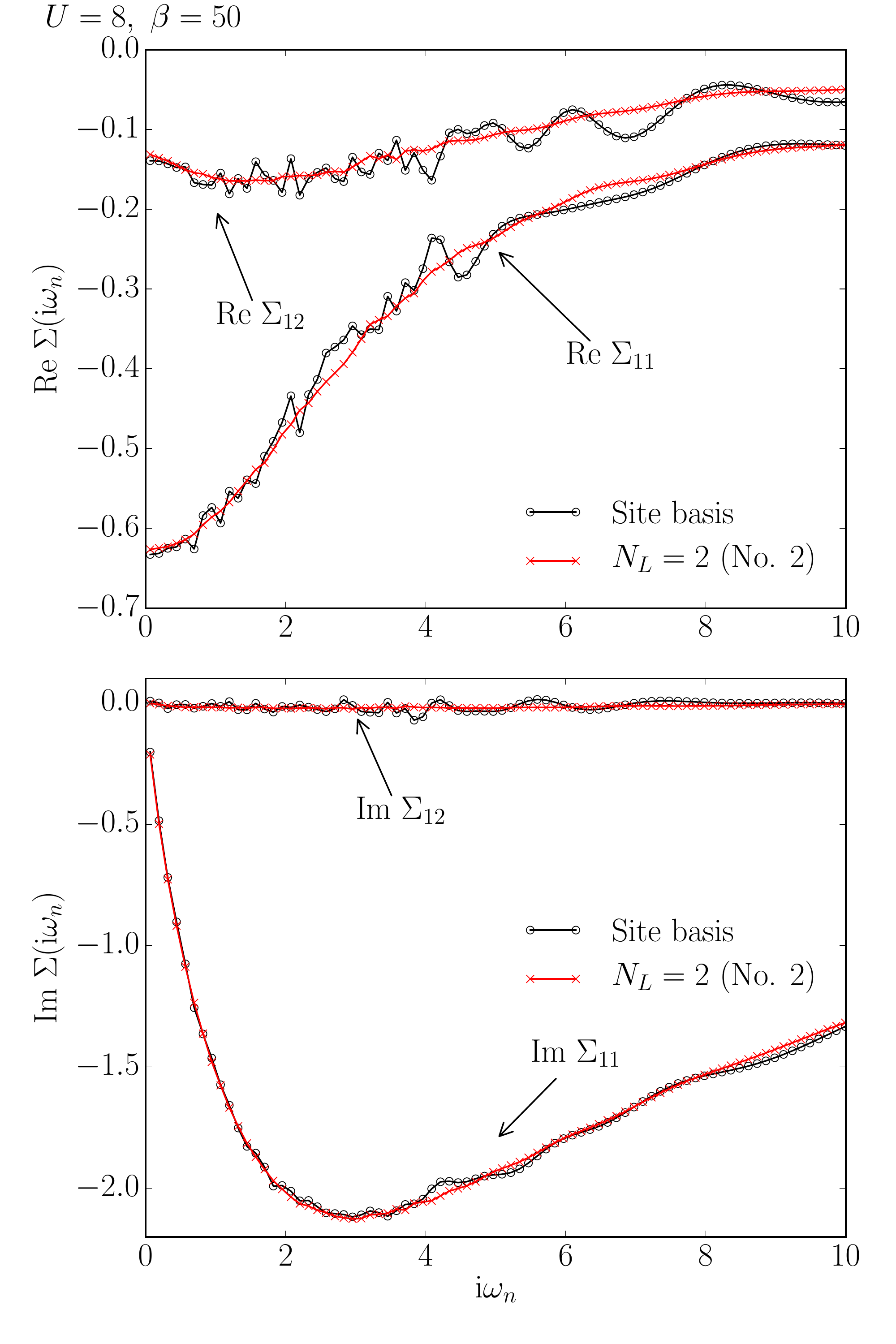}
	\caption{
		(Color online)
		Comparison of the self-energy $\Sigma(\mathrm{i}\omega_n)$ computed by the site basis and the $N_L=2$ (No. 2) basis using CT-INT at $U=8$ and $\beta=50$.
		We plot the site-diagonal element $\Sigma_{11}$ and the site-off-diagonal element $\Sigma_{12}$ in the site basis.
	}
	\label{fig:tetramer-ctint-sigma}
\end{figure}

\section{Summary and discussion}
In this paper, we showed that one can substantially reduce the negative sign problem in QMC simulations of quantum impurity models by using single-particle bases obtained by diagonalizing a part of the intracluster single-particle Hamiltonian.
First, we investigated the trimer and the tetramer using CT-HYB.
We found that optimal bases can be generated by diagonalizing subsets of the intracluster single-particle Hamiltonian in the impurity.
In particular, we revealed that single-particle bases consisting of bonding and anti-bonding orbitals have a high average sign when the kinetic energy and the onsite repulsion compete.
Furthermore, we tested these bases in the framework of CT-INT for the trimer and the tetramer models.
We showed that the sign problem is substantially suppressed with these bases, especially in the strongly correlated regime.

In the present study, we restricted our consideration to orbital-diagonal baths to focus on the negative sign problem arising from the local part of the impurity.
Our new bases may be useful when one applies cluster extensions of DMFT to Hubbard models on frustrated lattices such as a Kagom\'{e} lattice~\cite{Kita:2013ij,Udagawa:2010gj,Ohashi:2006kp,Syozi:1951ip,Ohashi:2008di,Mekata:2003ix} and a pyrochlore lattice.~\cite{Go:2012fb} 
On the other hand, for CT-HYB, off-diagonal elements of the hybridization function also give rise to a sign problem.
It remains to be clarified how to choose the optimal single-particle bases for impurity models with large off-diagonal elements. 
This will be practically important for investigating square-lattice and cubic-lattice Hubbard models using  cluster extensions of DMFT.

For CT-HYB, the local trace in Eq.~(\ref{eq:weight}) is evaluated either by the matrix formalism or by the Krylov formalism.~\cite{Lauchli:2009er}
Although we employed the matrix formalism in the present study,
the single-particle basis transformation applies to the Krylov formalism as well.
The Krylov algorithm may be more efficient than the matrix formalism when treating more than five orbitals.

For CT-INT, we restricted our consideration to single-particle bases consisting of independent pairs of bonding/anti-bonding orbitals.
The sign problem may be further reduced for more general single-particle bases which mix more than two sites.
Such basis transformations may generate general two-body interaction terms
$V_{ijkl} d^\dagger_{i\sigma} d^\dagger_{j\sigma^\prime} d_{k\sigma^\prime} d_{l\sigma}$ with $i\neq j \neq k \neq l$ ($i$, $j$, $k$, and $l$ are orbitals in the rotated basis) such as correlated hoppings.
One can still treat these general two-body interactions within CT-INT.~\cite{Gorelov:2009bi}

Finding the best single-particle basis is essentially a classical optimization problem.
Unitary matrices $\boldsymbol{U}$ can be parameterized as $\boldsymbol{U}=e^{\mathi \boldsymbol{H}}$, where $\boldsymbol{H}$ is a Hermitian matrix.~\footnote{
Orthogonal matrices are parameterized as $\boldsymbol{U} = e^{\boldsymbol{T}}$, 
where $\boldsymbol{T}$ is an anti-symmetric real matrix ($T_{ij}=-T_{ji}$).~\cite{Culver:1966bj,Sherif:2008vu}
}
We may be able to solve this optimization problem numerically, e.g., by using simulated annealing.~\cite{Kirkpatrick:1983zz,Cerny:1985bg}
Alternatively, constructions inspired from genetic algorithms might
be useful: in such a ``genetic'' scheme, basis sets for smaller clusters 
might be used as ``genes'' from which trial basis sets for larger clusters  
are constructed. Mutations would then correspond to adding or removing
links in the diagonalization process, and selection would be based on
the average sign as the fitness function. In this way, one might hope
to extend the present insights to clusters of much larger size
than in the present study.

\begin{acknowledgments}
We thank L. Boehnke, E. Brown, M. Harland, L. Huang, M. Iazzi, A. Lichtenstein, Y. Motome, H. Strand, L. Wang, N. Tsuji for stimulating discussions and useful comments.
We acknowledge support from the DFG via FOR 1346, the SNF Grant 200021E-149122, ERC Advanced Grant SIMCOFE, NCCR MARVEL and ERC Consolidator Grant CORRELMAT (project number 617196).
The calculations have been performed on the M\"{o}nch and Brutus clusters of ETH Z\"{u}rich using codes based on ALPS.~\cite{Bauer:2011tz}
\end{acknowledgments}

\bibliography{ref}

\begin{thebibliography}{49}%
\makeatletter
\providecommand \@ifxundefined [1]{%
 \@ifx{#1\undefined}
}%
\providecommand \@ifnum [1]{%
 \ifnum #1\expandafter \@firstoftwo
 \else \expandafter \@secondoftwo
 \fi
}%
\providecommand \@ifx [1]{%
 \ifx #1\expandafter \@firstoftwo
 \else \expandafter \@secondoftwo
 \fi
}%
\providecommand \natexlab [1]{#1}%
\providecommand \enquote  [1]{``#1''}%
\providecommand \bibnamefont  [1]{#1}%
\providecommand \bibfnamefont [1]{#1}%
\providecommand \citenamefont [1]{#1}%
\providecommand \href@noop [0]{\@secondoftwo}%
\providecommand \href [0]{\begingroup \@sanitize@url \@href}%
\providecommand \@href[1]{\@@startlink{#1}\@@href}%
\providecommand \@@href[1]{\endgroup#1\@@endlink}%
\providecommand \@sanitize@url [0]{\catcode `\\12\catcode `\$12\catcode
  `\&12\catcode `\#12\catcode `\^12\catcode `\_12\catcode `\%12\relax}%
\providecommand \@@startlink[1]{}%
\providecommand \@@endlink[0]{}%
\providecommand \url  [0]{\begingroup\@sanitize@url \@url }%
\providecommand \@url [1]{\endgroup\@href {#1}{\urlprefix }}%
\providecommand \urlprefix  [0]{URL }%
\providecommand \Eprint [0]{\href }%
\providecommand \doibase [0]{http://dx.doi.org/}%
\providecommand \selectlanguage [0]{\@gobble}%
\providecommand \bibinfo  [0]{\@secondoftwo}%
\providecommand \bibfield  [0]{\@secondoftwo}%
\providecommand \translation [1]{[#1]}%
\providecommand \BibitemOpen [0]{}%
\providecommand \bibitemStop [0]{}%
\providecommand \bibitemNoStop [0]{.\EOS\space}%
\providecommand \EOS [0]{\spacefactor3000\relax}%
\providecommand \BibitemShut  [1]{\csname bibitem#1\endcsname}%
\let\auto@bib@innerbib\@empty
%</preamble>
\bibitem [{\citenamefont {Loh~Jr}\ \emph {et~al.}(1990)\citenamefont {Loh~Jr},
  \citenamefont {Gubernatis}, \citenamefont {Scalettar},\ and\ \citenamefont
  {White}}]{LohJr:1990up}%
  \BibitemOpen
  \bibfield  {author} {\bibinfo {author} {\bibfnamefont {E.~Y.}\ \bibnamefont
  {Loh~Jr}}, \bibinfo {author} {\bibfnamefont {J.~E.}\ \bibnamefont
  {Gubernatis}}, \bibinfo {author} {\bibfnamefont {R.~T.}\ \bibnamefont
  {Scalettar}}, \ and\ \bibinfo {author} {\bibfnamefont {S.~R.}\ \bibnamefont
  {White}},\ }\href
  {http://journals.aps.org/prb/abstract/10.1103/PhysRevB.41.9301} {\bibfield
  {journal} {\bibinfo  {journal} {Physical Review B}\ }\textbf {\bibinfo
  {volume} {41}},\ \bibinfo {pages} {9301} (\bibinfo {year}
  {1990})}\BibitemShut {NoStop}%
\bibitem [{\citenamefont {Hirsch}(1985)}]{Hirsch:1985wz}%
  \BibitemOpen
  \bibfield  {author} {\bibinfo {author} {\bibfnamefont {J.~E.}\ \bibnamefont
  {Hirsch}},\ }\href
  {http://gateway.webofknowledge.com/gateway/Gateway.cgi?GWVersion=2&SrcAuth=m%
ekentosj&SrcApp=Papers&DestLinkType=FullRecord&DestApp=WOS&KeyUT=A1985AEY85000%
37} {\bibfield  {journal} {\bibinfo  {journal} {Physical Review B}\ }\textbf
  {\bibinfo {volume} {31}},\ \bibinfo {pages} {4403} (\bibinfo {year}
  {1985})}\BibitemShut {NoStop}%
\bibitem [{\citenamefont {Chandrasekharan}\ and\ \citenamefont
  {Wiese}(1999)}]{Chandrasekharan:1999kn}%
  \BibitemOpen
  \bibfield  {author} {\bibinfo {author} {\bibfnamefont {S.}~\bibnamefont
  {Chandrasekharan}}\ and\ \bibinfo {author} {\bibfnamefont {U.~J.}\
  \bibnamefont {Wiese}},\ }\href {\doibase 10.1103/PhysRevLett.83.3116}
  {\bibfield  {journal} {\bibinfo  {journal} {Physical Review Letters}\
  }\textbf {\bibinfo {volume} {83}},\ \bibinfo {pages} {3116} (\bibinfo {year}
  {1999})}\BibitemShut {NoStop}%
\bibitem [{\citenamefont {Capponi}\ and\ \citenamefont
  {Assaad}(2001)}]{Capponi:2001jc}%
  \BibitemOpen
  \bibfield  {author} {\bibinfo {author} {\bibfnamefont {S.}~\bibnamefont
  {Capponi}}\ and\ \bibinfo {author} {\bibfnamefont {F.~F.}\ \bibnamefont
  {Assaad}},\ }\href {\doibase 10.1103/PhysRevB.63.155114} {\bibfield
  {journal} {\bibinfo  {journal} {Physical Review B}\ }\textbf {\bibinfo
  {volume} {63}},\ \bibinfo {pages} {155114} (\bibinfo {year}
  {2001})}\BibitemShut {NoStop}%
\bibitem [{\citenamefont {Wu}\ \emph {et~al.}(2003)\citenamefont {Wu},
  \citenamefont {Hu},\ and\ \citenamefont {Zhang}}]{Wu:2003es}%
  \BibitemOpen
  \bibfield  {author} {\bibinfo {author} {\bibfnamefont {C.}~\bibnamefont
  {Wu}}, \bibinfo {author} {\bibfnamefont {J.-p.}\ \bibnamefont {Hu}}, \ and\
  \bibinfo {author} {\bibfnamefont {S.-C.}\ \bibnamefont {Zhang}},\ }\href
  {\doibase 10.1103/PhysRevLett.91.186402} {\bibfield  {journal} {\bibinfo
  {journal} {Physical Review Letters}\ }\textbf {\bibinfo {volume} {91}},\
  \bibinfo {pages} {186402} (\bibinfo {year} {2003})}\BibitemShut {NoStop}%
\bibitem [{\citenamefont {Wu}\ and\ \citenamefont {Zhang}(2005)}]{Wu:2005im}%
  \BibitemOpen
  \bibfield  {author} {\bibinfo {author} {\bibfnamefont {C.}~\bibnamefont
  {Wu}}\ and\ \bibinfo {author} {\bibfnamefont {S.-C.}\ \bibnamefont {Zhang}},\
  }\href {\doibase 10.1103/PhysRevB.71.155115} {\bibfield  {journal} {\bibinfo
  {journal} {Physical Review B}\ }\textbf {\bibinfo {volume} {71}},\ \bibinfo
  {pages} {155115} (\bibinfo {year} {2005})}\BibitemShut {NoStop}%
\bibitem [{\citenamefont {Li}\ \emph {et~al.}(2015{\natexlab{a}})\citenamefont
  {Li}, \citenamefont {Jiang},\ and\ \citenamefont {Yao}}]{Li:2015cw}%
  \BibitemOpen
  \bibfield  {author} {\bibinfo {author} {\bibfnamefont {Z.-X.}\ \bibnamefont
  {Li}}, \bibinfo {author} {\bibfnamefont {Y.-F.}\ \bibnamefont {Jiang}}, \
  and\ \bibinfo {author} {\bibfnamefont {H.}~\bibnamefont {Yao}},\ }\href
  {\doibase 10.1088/1367-2630/17/8/085003} {\bibfield  {journal} {\bibinfo
  {journal} {New Journal of Physics}\ }\textbf {\bibinfo {volume} {17}},\
  \bibinfo {pages} {1} (\bibinfo {year} {2015}{\natexlab{a}})}\BibitemShut
  {NoStop}%
\bibitem [{\citenamefont {Li}\ \emph {et~al.}(2015{\natexlab{b}})\citenamefont
  {Li}, \citenamefont {Jiang},\ and\ \citenamefont {Yao}}]{Li:2015jf}%
  \BibitemOpen
  \bibfield  {author} {\bibinfo {author} {\bibfnamefont {Z.-X.}\ \bibnamefont
  {Li}}, \bibinfo {author} {\bibfnamefont {Y.-F.}\ \bibnamefont {Jiang}}, \
  and\ \bibinfo {author} {\bibfnamefont {H.}~\bibnamefont {Yao}},\ }\href
  {\doibase 10.1103/PhysRevB.91.241117} {\bibfield  {journal} {\bibinfo
  {journal} {Physical Review B}\ }\textbf {\bibinfo {volume} {91}},\ \bibinfo
  {pages} {241117} (\bibinfo {year} {2015}{\natexlab{b}})}\BibitemShut
  {NoStop}%
\bibitem [{\citenamefont {Wang}\ \emph {et~al.}(2015)\citenamefont {Wang},
  \citenamefont {Liu}, \citenamefont {Iazzi}, \citenamefont {Troyer},\ and\
  \citenamefont {Harcos}}]{Wang:2015vha}%
  \BibitemOpen
  \bibfield  {author} {\bibinfo {author} {\bibfnamefont {L.}~\bibnamefont
  {Wang}}, \bibinfo {author} {\bibfnamefont {Y.-H.}\ \bibnamefont {Liu}},
  \bibinfo {author} {\bibfnamefont {M.}~\bibnamefont {Iazzi}}, \bibinfo
  {author} {\bibfnamefont {M.}~\bibnamefont {Troyer}}, \ and\ \bibinfo {author}
  {\bibfnamefont {G.}~\bibnamefont {Harcos}},\ }\href
  {http://arxiv.org/abs/1506.05349v2} {\bibfield  {journal} {\bibinfo
  {journal} {Physical Review B}\ } (\bibinfo {year} {2015})},\ \Eprint
  {http://arxiv.org/abs/1506.05349v2} {1506.05349v2} \BibitemShut {NoStop}%
\bibitem [{\citenamefont {Ferrero}\ \emph {et~al.}(2007)\citenamefont
  {Ferrero}, \citenamefont {De~Leo}, \citenamefont {Lecheminant},\ and\
  \citenamefont {Fabrizio}}]{Ferrero:2007bz}%
  \BibitemOpen
  \bibfield  {author} {\bibinfo {author} {\bibfnamefont {M.}~\bibnamefont
  {Ferrero}}, \bibinfo {author} {\bibfnamefont {L.}~\bibnamefont {De~Leo}},
  \bibinfo {author} {\bibfnamefont {P.}~\bibnamefont {Lecheminant}}, \ and\
  \bibinfo {author} {\bibfnamefont {M.}~\bibnamefont {Fabrizio}},\ }\href
  {\doibase 10.1088/0953-8984/19/43/433201} {\bibfield  {journal} {\bibinfo
  {journal} {Journal of Physics: Condensed Matter}\ }\textbf {\bibinfo {volume}
  {19}},\ \bibinfo {pages} {433201} (\bibinfo {year} {2007})}\BibitemShut
  {NoStop}%
\bibitem [{\citenamefont {Anderson}(1961)}]{Anderson:1961}%
  \BibitemOpen
  \bibfield  {author} {\bibinfo {author} {\bibfnamefont {P.~W.}\ \bibnamefont
  {Anderson}},\ }\href
  {http://journals.aps.org/pr/abstract/10.1103/PhysRev.124.41} {\bibfield
  {journal} {\bibinfo  {journal} {Physical Review}\ }\textbf {\bibinfo {volume}
  {124}},\ \bibinfo {pages} {41} (\bibinfo {year} {1961})}\BibitemShut
  {NoStop}%
\bibitem [{\citenamefont {Hewson}(1997)}]{Hewson:1997vc}%
  \BibitemOpen
  \bibfield  {author} {\bibinfo {author} {\bibfnamefont {A.~C.}\ \bibnamefont
  {Hewson}},\ }\href {http://adsabs.harvard.edu/abs/1997kphf.book.....C} {\emph
  {\bibinfo {title} {{The Kondo Problem to Heavy Fermions}}}}\ (\bibinfo
  {publisher} {Cambridge University Press},\ \bibinfo {year}
  {1997})\BibitemShut {NoStop}%
\bibitem [{\citenamefont {Van~Dongen}\ and\ \citenamefont
  {Vollhardt}(1990)}]{VanDongen:1990df}%
  \BibitemOpen
  \bibfield  {author} {\bibinfo {author} {\bibfnamefont {P.}~\bibnamefont
  {Van~Dongen}}\ and\ \bibinfo {author} {\bibfnamefont {D.}~\bibnamefont
  {Vollhardt}},\ }\href {\doibase 10.1103/PhysRevLett.65.1663} {\bibfield
  {journal} {\bibinfo  {journal} {Physical Review Letters}\ }\textbf {\bibinfo
  {volume} {65}},\ \bibinfo {pages} {1663} (\bibinfo {year}
  {1990})}\BibitemShut {NoStop}%
\bibitem [{\citenamefont {Metzner}\ and\ \citenamefont
  {Vollhardt}(1989)}]{Metzner:1989bz}%
  \BibitemOpen
  \bibfield  {author} {\bibinfo {author} {\bibfnamefont {W.}~\bibnamefont
  {Metzner}}\ and\ \bibinfo {author} {\bibfnamefont {D.}~\bibnamefont
  {Vollhardt}},\ }\href {\doibase 10.1103/PhysRevLett.62.324} {\bibfield
  {journal} {\bibinfo  {journal} {Physical Review Letters}\ }\textbf {\bibinfo
  {volume} {62}},\ \bibinfo {pages} {324} (\bibinfo {year} {1989})}\BibitemShut
  {NoStop}%
\bibitem [{\citenamefont {Georges}\ \emph {et~al.}(1996)\citenamefont
  {Georges}, \citenamefont {Kotliar}, \citenamefont {Krauth},\ and\
  \citenamefont {Rozenberg}}]{Georges:1996un}%
  \BibitemOpen
  \bibfield  {author} {\bibinfo {author} {\bibfnamefont {A.}~\bibnamefont
  {Georges}}, \bibinfo {author} {\bibfnamefont {G.}~\bibnamefont {Kotliar}},
  \bibinfo {author} {\bibfnamefont {W.}~\bibnamefont {Krauth}}, \ and\ \bibinfo
  {author} {\bibfnamefont {M.~J.}\ \bibnamefont {Rozenberg}},\ }\href@noop {}
  {\bibfield  {journal} {\bibinfo  {journal} {Reviews of Modern Physics}\
  }\textbf {\bibinfo {volume} {68}},\ \bibinfo {pages} {13} (\bibinfo {year}
  {1996})}\BibitemShut {NoStop}%
\bibitem [{\citenamefont {Maier}\ \emph {et~al.}(2005)\citenamefont {Maier},
  \citenamefont {Jarrell}, \citenamefont {Pruschke},\ and\ \citenamefont
  {Hettler}}]{Maier:2005et}%
  \BibitemOpen
  \bibfield  {author} {\bibinfo {author} {\bibfnamefont {T.}~\bibnamefont
  {Maier}}, \bibinfo {author} {\bibfnamefont {M.}~\bibnamefont {Jarrell}},
  \bibinfo {author} {\bibfnamefont {T.}~\bibnamefont {Pruschke}}, \ and\
  \bibinfo {author} {\bibfnamefont {M.~H.}\ \bibnamefont {Hettler}},\ }\href
  {\doibase 10.1103/RevModPhys.77.1027} {\bibfield  {journal} {\bibinfo
  {journal} {Reviews of Modern Physics}\ }\textbf {\bibinfo {volume} {77}},\
  \bibinfo {pages} {1027} (\bibinfo {year} {2005})}\BibitemShut {NoStop}%
\bibitem [{\citenamefont {Lichtenstein}\ and\ \citenamefont
  {Katsnelson}(2000)}]{Lichtenstein:2000bp}%
  \BibitemOpen
  \bibfield  {author} {\bibinfo {author} {\bibfnamefont {A.~I.}\ \bibnamefont
  {Lichtenstein}}\ and\ \bibinfo {author} {\bibfnamefont {M.~I.}\ \bibnamefont
  {Katsnelson}},\ }\href {\doibase 10.1103/PhysRevB.62.R9283} {\bibfield
  {journal} {\bibinfo  {journal} {Physical Review B}\ }\textbf {\bibinfo
  {volume} {62}},\ \bibinfo {pages} {R9283} (\bibinfo {year}
  {2000})}\BibitemShut {NoStop}%
\bibitem [{\citenamefont {Kotliar}\ \emph {et~al.}(2001)\citenamefont
  {Kotliar}, \citenamefont {Savrasov}, \citenamefont {P{\'a}lsson},\ and\
  \citenamefont {Biroli}}]{Kotliar:2001iy}%
  \BibitemOpen
  \bibfield  {author} {\bibinfo {author} {\bibfnamefont {G.}~\bibnamefont
  {Kotliar}}, \bibinfo {author} {\bibfnamefont {S.}~\bibnamefont {Savrasov}},
  \bibinfo {author} {\bibfnamefont {G.}~\bibnamefont {P{\'a}lsson}}, \ and\
  \bibinfo {author} {\bibfnamefont {G.}~\bibnamefont {Biroli}},\ }\href
  {\doibase 10.1103/PhysRevLett.87.186401} {\bibfield  {journal} {\bibinfo
  {journal} {Physical Review Letters}\ }\textbf {\bibinfo {volume} {87}},\
  \bibinfo {pages} {186401} (\bibinfo {year} {2001})}\BibitemShut {NoStop}%
\bibitem [{\citenamefont {Hettler}\ \emph {et~al.}(2000)\citenamefont
  {Hettler}, \citenamefont {Mukherjee}, \citenamefont {Jarrell},\ and\
  \citenamefont {Krishnamurthy}}]{Hettler:2000es}%
  \BibitemOpen
  \bibfield  {author} {\bibinfo {author} {\bibfnamefont {M.~H.}\ \bibnamefont
  {Hettler}}, \bibinfo {author} {\bibfnamefont {M.}~\bibnamefont {Mukherjee}},
  \bibinfo {author} {\bibfnamefont {M.}~\bibnamefont {Jarrell}}, \ and\
  \bibinfo {author} {\bibfnamefont {H.~R.}\ \bibnamefont {Krishnamurthy}},\
  }\href {\doibase 10.1103/PhysRevB.61.12739} {\bibfield  {journal} {\bibinfo
  {journal} {Physical Review B}\ }\textbf {\bibinfo {volume} {61}},\ \bibinfo
  {pages} {12739} (\bibinfo {year} {2000})}\BibitemShut {NoStop}%
\bibitem [{\citenamefont {Rubtsov}\ \emph {et~al.}(2005)\citenamefont
  {Rubtsov}, \citenamefont {Savkin},\ and\ \citenamefont
  {Lichtenstein}}]{Rubtsov:2005iwa}%
  \BibitemOpen
  \bibfield  {author} {\bibinfo {author} {\bibfnamefont {A.}~\bibnamefont
  {Rubtsov}}, \bibinfo {author} {\bibfnamefont {V.}~\bibnamefont {Savkin}}, \
  and\ \bibinfo {author} {\bibfnamefont {A.}~\bibnamefont {Lichtenstein}},\
  }\href {\doibase 10.1103/PhysRevB.72.035122} {\bibfield  {journal} {\bibinfo
  {journal} {Physical Review B}\ }\textbf {\bibinfo {volume} {72}},\ \bibinfo
  {pages} {035122} (\bibinfo {year} {2005})}\BibitemShut {NoStop}%
\bibitem [{\citenamefont {Gull}\ \emph {et~al.}(2008)\citenamefont {Gull},
  \citenamefont {Werner}, \citenamefont {Parcollet},\ and\ \citenamefont
  {Troyer}}]{Gull:2008cma}%
  \BibitemOpen
  \bibfield  {author} {\bibinfo {author} {\bibfnamefont {E.}~\bibnamefont
  {Gull}}, \bibinfo {author} {\bibfnamefont {P.}~\bibnamefont {Werner}},
  \bibinfo {author} {\bibfnamefont {O.}~\bibnamefont {Parcollet}}, \ and\
  \bibinfo {author} {\bibfnamefont {M.}~\bibnamefont {Troyer}},\ }\href
  {\doibase 10.1209/0295-5075/82/57003} {\bibfield  {journal} {\bibinfo
  {journal} {EPL (Europhysics Letters)}\ }\textbf {\bibinfo {volume} {82}},\
  \bibinfo {pages} {57003} (\bibinfo {year} {2008})}\BibitemShut {NoStop}%
\bibitem [{\citenamefont {Werner}\ \emph {et~al.}(2006)\citenamefont {Werner},
  \citenamefont {Comanac}, \citenamefont {de{\textquoteright} Medici},
  \citenamefont {Troyer},\ and\ \citenamefont {Millis}}]{Werner:2006ko}%
  \BibitemOpen
  \bibfield  {author} {\bibinfo {author} {\bibfnamefont {P.}~\bibnamefont
  {Werner}}, \bibinfo {author} {\bibfnamefont {A.}~\bibnamefont {Comanac}},
  \bibinfo {author} {\bibfnamefont {L.}~\bibnamefont {de{\textquoteright}
  Medici}}, \bibinfo {author} {\bibfnamefont {M.}~\bibnamefont {Troyer}}, \
  and\ \bibinfo {author} {\bibfnamefont {A.}~\bibnamefont {Millis}},\ }\href
  {\doibase 10.1103/PhysRevLett.97.076405} {\bibfield  {journal} {\bibinfo
  {journal} {Physical Review Letters}\ }\textbf {\bibinfo {volume} {97}},\
  \bibinfo {pages} {076405} (\bibinfo {year} {2006})}\BibitemShut {NoStop}%
\bibitem [{\citenamefont {Werner}\ and\ \citenamefont
  {Millis}(2006)}]{Werner:2006iz}%
  \BibitemOpen
  \bibfield  {author} {\bibinfo {author} {\bibfnamefont {P.}~\bibnamefont
  {Werner}}\ and\ \bibinfo {author} {\bibfnamefont {A.}~\bibnamefont
  {Millis}},\ }\href {\doibase 10.1103/PhysRevB.74.155107} {\bibfield
  {journal} {\bibinfo  {journal} {Physical Review B}\ }\textbf {\bibinfo
  {volume} {74}},\ \bibinfo {pages} {155107} (\bibinfo {year}
  {2006})}\BibitemShut {NoStop}%
\bibitem [{\citenamefont {Rubtsov}(2003)}]{Rubtsov:2003vb}%
  \BibitemOpen
  \bibfield  {author} {\bibinfo {author} {\bibfnamefont {A.~N.}\ \bibnamefont
  {Rubtsov}},\ }\href {http://arxiv.org/abs/cond-mat/0302228v1} {\bibfield
  {journal} {\bibinfo  {journal} {Physical Review B}\ } (\bibinfo {year}
  {2003})},\ \Eprint {http://arxiv.org/abs/cond-mat/0302228v1}
  {cond-mat/0302228v1} \BibitemShut {NoStop}%
\bibitem [{\citenamefont {Yoo}\ \emph {et~al.}(2005)\citenamefont {Yoo},
  \citenamefont {Chandrasekharan}, \citenamefont {Kaul}, \citenamefont
  {Ullmo},\ and\ \citenamefont {Baranger}}]{Yoo:2005ib}%
  \BibitemOpen
  \bibfield  {author} {\bibinfo {author} {\bibfnamefont {J.}~\bibnamefont
  {Yoo}}, \bibinfo {author} {\bibfnamefont {S.}~\bibnamefont
  {Chandrasekharan}}, \bibinfo {author} {\bibfnamefont {R.~K.}\ \bibnamefont
  {Kaul}}, \bibinfo {author} {\bibfnamefont {D.}~\bibnamefont {Ullmo}}, \ and\
  \bibinfo {author} {\bibfnamefont {H.~U.}\ \bibnamefont {Baranger}},\ }\href
  {\doibase 10.1088/0305-4470/38/48/004} {\bibfield  {journal} {\bibinfo
  {journal} {Journal of Physics A: Mathematical and General}\ }\textbf
  {\bibinfo {volume} {38}},\ \bibinfo {pages} {10307} (\bibinfo {year}
  {2005})}\BibitemShut {NoStop}%
\bibitem [{\citenamefont {Furukawa}\ \emph {et~al.}(2010)\citenamefont
  {Furukawa}, \citenamefont {Ohashi}, \citenamefont {Koyama},\ and\
  \citenamefont {Kawakami}}]{Furukawa:2010bx}%
  \BibitemOpen
  \bibfield  {author} {\bibinfo {author} {\bibfnamefont {Y.}~\bibnamefont
  {Furukawa}}, \bibinfo {author} {\bibfnamefont {T.}~\bibnamefont {Ohashi}},
  \bibinfo {author} {\bibfnamefont {Y.}~\bibnamefont {Koyama}}, \ and\ \bibinfo
  {author} {\bibfnamefont {N.}~\bibnamefont {Kawakami}},\ }\href {\doibase
  10.1103/PhysRevB.82.161101} {\bibfield  {journal} {\bibinfo  {journal}
  {Physical Review B}\ }\textbf {\bibinfo {volume} {82}},\ \bibinfo {pages}
  {161101} (\bibinfo {year} {2010})}\BibitemShut {NoStop}%
\bibitem [{\citenamefont {Sato}\ \emph {et~al.}(2015)\citenamefont {Sato},
  \citenamefont {Shirakawa},\ and\ \citenamefont
  {Yunoki}}]{Anonymous:5bgqkPSw}%
  \BibitemOpen
  \bibfield  {author} {\bibinfo {author} {\bibfnamefont {T.}~\bibnamefont
  {Sato}}, \bibinfo {author} {\bibfnamefont {T.}~\bibnamefont {Shirakawa}}, \
  and\ \bibinfo {author} {\bibfnamefont {S.}~\bibnamefont {Yunoki}},\ }\href
  {http://arxiv.org/abs/1502.00108v1} {\bibfield  {journal} {\bibinfo
  {journal} {Physical Review B}\ } (\bibinfo {year} {2015})},\ \Eprint
  {http://arxiv.org/abs/1502.00108v1} {1502.00108v1} \BibitemShut {NoStop}%
\bibitem [{Note1()}]{Note1}%
  \BibitemOpen
  \bibinfo {note} {We used the numpy package for Python.}\BibitemShut {Stop}%
\bibitem [{\citenamefont {Hattori}\ and\ \citenamefont
  {Tsunetsugu}(2012)}]{Hattori:2012bh}%
  \BibitemOpen
  \bibfield  {author} {\bibinfo {author} {\bibfnamefont {K.}~\bibnamefont
  {Hattori}}\ and\ \bibinfo {author} {\bibfnamefont {H.}~\bibnamefont
  {Tsunetsugu}},\ }\href {\doibase 10.1103/PhysRevB.86.054421} {\bibfield
  {journal} {\bibinfo  {journal} {Physical Review B}\ }\textbf {\bibinfo
  {volume} {86}},\ \bibinfo {pages} {054421} (\bibinfo {year}
  {2012})}\BibitemShut {NoStop}%
\bibitem [{Note2()}]{Note2}%
  \BibitemOpen
  \bibinfo {note} {We confirmed that $A_1$ ($T_2$) indeed has the highest
  average sign for $0.2\le t\le 1$ and $\beta =50$ among transformation
  matrices which diagonalize the intracluster single-particle Hamiltonian of
  $N_L=6$.}\BibitemShut {Stop}%
\bibitem [{\citenamefont {Gull}\ \emph {et~al.}(2011)\citenamefont {Gull},
  \citenamefont {Millis}, \citenamefont {Lichtenstein}, \citenamefont
  {Rubtsov}, \citenamefont {Troyer},\ and\ \citenamefont
  {Werner}}]{Gull:2011jda}%
  \BibitemOpen
  \bibfield  {author} {\bibinfo {author} {\bibfnamefont {E.}~\bibnamefont
  {Gull}}, \bibinfo {author} {\bibfnamefont {A.~J.}\ \bibnamefont {Millis}},
  \bibinfo {author} {\bibfnamefont {A.~I.}\ \bibnamefont {Lichtenstein}},
  \bibinfo {author} {\bibfnamefont {A.~N.}\ \bibnamefont {Rubtsov}}, \bibinfo
  {author} {\bibfnamefont {M.}~\bibnamefont {Troyer}}, \ and\ \bibinfo {author}
  {\bibfnamefont {P.}~\bibnamefont {Werner}},\ }\href {\doibase
  10.1103/RevModPhys.83.349} {\bibfield  {journal} {\bibinfo  {journal}
  {Reviews of Modern Physics}\ }\textbf {\bibinfo {volume} {83}},\ \bibinfo
  {pages} {349} (\bibinfo {year} {2011})}\BibitemShut {NoStop}%
\bibitem [{\citenamefont {Nomura}\ \emph {et~al.}(2014)\citenamefont {Nomura},
  \citenamefont {Sakai},\ and\ \citenamefont {Arita}}]{Nomura:2014by}%
  \BibitemOpen
  \bibfield  {author} {\bibinfo {author} {\bibfnamefont {Y.}~\bibnamefont
  {Nomura}}, \bibinfo {author} {\bibfnamefont {S.}~\bibnamefont {Sakai}}, \
  and\ \bibinfo {author} {\bibfnamefont {R.}~\bibnamefont {Arita}},\ }\href
  {\doibase 10.1103/PhysRevB.89.195146} {\bibfield  {journal} {\bibinfo
  {journal} {Physical Review B}\ }\textbf {\bibinfo {volume} {89}},\ \bibinfo
  {pages} {195146} (\bibinfo {year} {2014})}\BibitemShut {NoStop}%
\bibitem [{\citenamefont {Nomura}\ \emph {et~al.}(2015)\citenamefont {Nomura},
  \citenamefont {Sakai},\ and\ \citenamefont {Arita}}]{Nomura:2015cb}%
  \BibitemOpen
  \bibfield  {author} {\bibinfo {author} {\bibfnamefont {Y.}~\bibnamefont
  {Nomura}}, \bibinfo {author} {\bibfnamefont {S.}~\bibnamefont {Sakai}}, \
  and\ \bibinfo {author} {\bibfnamefont {R.}~\bibnamefont {Arita}},\ }\href
  {\doibase 10.1103/PhysRevB.91.235107} {\bibfield  {journal} {\bibinfo
  {journal} {Physical Review B}\ }\textbf {\bibinfo {volume} {91}},\ \bibinfo
  {pages} {235107} (\bibinfo {year} {2015})}\BibitemShut {NoStop}%
\bibitem [{\citenamefont {Boehnke}\ \emph {et~al.}(2011)\citenamefont
  {Boehnke}, \citenamefont {Hafermann}, \citenamefont {Ferrero}, \citenamefont
  {Lechermann},\ and\ \citenamefont {Parcollet}}]{Boehnke:2011dd}%
  \BibitemOpen
  \bibfield  {author} {\bibinfo {author} {\bibfnamefont {L.}~\bibnamefont
  {Boehnke}}, \bibinfo {author} {\bibfnamefont {H.}~\bibnamefont {Hafermann}},
  \bibinfo {author} {\bibfnamefont {M.}~\bibnamefont {Ferrero}}, \bibinfo
  {author} {\bibfnamefont {F.}~\bibnamefont {Lechermann}}, \ and\ \bibinfo
  {author} {\bibfnamefont {O.}~\bibnamefont {Parcollet}},\ }\href {\doibase
  10.1103/PhysRevB.84.075145} {\bibfield  {journal} {\bibinfo  {journal}
  {Physical Review B}\ }\textbf {\bibinfo {volume} {84}},\ \bibinfo {pages}
  {075145} (\bibinfo {year} {2011})}\BibitemShut {NoStop}%
\bibitem [{\citenamefont {Kita}\ \emph {et~al.}(2013)\citenamefont {Kita},
  \citenamefont {Ohashi},\ and\ \citenamefont {Kawakami}}]{Kita:2013ij}%
  \BibitemOpen
  \bibfield  {author} {\bibinfo {author} {\bibfnamefont {T.}~\bibnamefont
  {Kita}}, \bibinfo {author} {\bibfnamefont {T.}~\bibnamefont {Ohashi}}, \ and\
  \bibinfo {author} {\bibfnamefont {N.}~\bibnamefont {Kawakami}},\ }\href
  {\doibase 10.1103/PhysRevB.87.155119} {\bibfield  {journal} {\bibinfo
  {journal} {Physical Review B}\ }\textbf {\bibinfo {volume} {87}},\ \bibinfo
  {pages} {155119} (\bibinfo {year} {2013})}\BibitemShut {NoStop}%
\bibitem [{\citenamefont {Udagawa}\ and\ \citenamefont
  {Motome}(2010)}]{Udagawa:2010gj}%
  \BibitemOpen
  \bibfield  {author} {\bibinfo {author} {\bibfnamefont {M.}~\bibnamefont
  {Udagawa}}\ and\ \bibinfo {author} {\bibfnamefont {Y.}~\bibnamefont
  {Motome}},\ }\href {\doibase 10.1103/PhysRevLett.104.106409} {\bibfield
  {journal} {\bibinfo  {journal} {Physical Review Letters}\ }\textbf {\bibinfo
  {volume} {104}},\ \bibinfo {pages} {106409} (\bibinfo {year}
  {2010})}\BibitemShut {NoStop}%
\bibitem [{\citenamefont {Ohashi}\ \emph {et~al.}(2006)\citenamefont {Ohashi},
  \citenamefont {Kawakami},\ and\ \citenamefont {Tsunetsugu}}]{Ohashi:2006kp}%
  \BibitemOpen
  \bibfield  {author} {\bibinfo {author} {\bibfnamefont {T.}~\bibnamefont
  {Ohashi}}, \bibinfo {author} {\bibfnamefont {N.}~\bibnamefont {Kawakami}}, \
  and\ \bibinfo {author} {\bibfnamefont {H.}~\bibnamefont {Tsunetsugu}},\
  }\href {\doibase 10.1103/PhysRevLett.97.066401} {\bibfield  {journal}
  {\bibinfo  {journal} {Physical Review Letters}\ }\textbf {\bibinfo {volume}
  {97}},\ \bibinfo {pages} {066401} (\bibinfo {year} {2006})}\BibitemShut
  {NoStop}%
\bibitem [{\citenamefont {Sy{\^o}zi}(1951)}]{Syozi:1951ip}%
  \BibitemOpen
  \bibfield  {author} {\bibinfo {author} {\bibfnamefont {I.}~\bibnamefont
  {Sy{\^o}zi}},\ }\href {\doibase 10.1143/ptp/6.3.306} {\bibfield  {journal}
  {\bibinfo  {journal} {Progress of Theoretical Physics}\ }\textbf {\bibinfo
  {volume} {6}},\ \bibinfo {pages} {306} (\bibinfo {year} {1951})}\BibitemShut
  {NoStop}%
\bibitem [{\citenamefont {Ohashi}\ \emph {et~al.}(2008)\citenamefont {Ohashi},
  \citenamefont {Momoi}, \citenamefont {Tsunetsugu},\ and\ \citenamefont
  {Kawakami}}]{Ohashi:2008di}%
  \BibitemOpen
  \bibfield  {author} {\bibinfo {author} {\bibfnamefont {T.}~\bibnamefont
  {Ohashi}}, \bibinfo {author} {\bibfnamefont {T.}~\bibnamefont {Momoi}},
  \bibinfo {author} {\bibfnamefont {H.}~\bibnamefont {Tsunetsugu}}, \ and\
  \bibinfo {author} {\bibfnamefont {N.}~\bibnamefont {Kawakami}},\ }\href
  {\doibase 10.1103/PhysRevLett.100.076402} {\bibfield  {journal} {\bibinfo
  {journal} {Physical Review Letters}\ }\textbf {\bibinfo {volume} {100}},\
  \bibinfo {pages} {076402} (\bibinfo {year} {2008})}\BibitemShut {NoStop}%
\bibitem [{\citenamefont {Mekata}(2003)}]{Mekata:2003ix}%
  \BibitemOpen
  \bibfield  {author} {\bibinfo {author} {\bibfnamefont {M.}~\bibnamefont
  {Mekata}},\ }\href {\doibase 10.1063/1.1564329} {\bibfield  {journal}
  {\bibinfo  {journal} {Physics Today}\ }\textbf {\bibinfo {volume} {56}},\
  \bibinfo {pages} {12} (\bibinfo {year} {2003})}\BibitemShut {NoStop}%
\bibitem [{\citenamefont {Go}\ \emph {et~al.}(2012)\citenamefont {Go},
  \citenamefont {Witczak-Krempa}, \citenamefont {Jeon}, \citenamefont {Park},\
  and\ \citenamefont {Kim}}]{Go:2012fb}%
  \BibitemOpen
  \bibfield  {author} {\bibinfo {author} {\bibfnamefont {A.}~\bibnamefont
  {Go}}, \bibinfo {author} {\bibfnamefont {W.}~\bibnamefont {Witczak-Krempa}},
  \bibinfo {author} {\bibfnamefont {G.~S.}\ \bibnamefont {Jeon}}, \bibinfo
  {author} {\bibfnamefont {K.}~\bibnamefont {Park}}, \ and\ \bibinfo {author}
  {\bibfnamefont {Y.~B.}\ \bibnamefont {Kim}},\ }\href {\doibase
  10.1103/PhysRevLett.109.066401} {\bibfield  {journal} {\bibinfo  {journal}
  {Physical Review Letters}\ }\textbf {\bibinfo {volume} {109}},\ \bibinfo
  {pages} {066401} (\bibinfo {year} {2012})}\BibitemShut {NoStop}%
\bibitem [{\citenamefont {L{\"a}uchli}\ and\ \citenamefont
  {Werner}(2009)}]{Lauchli:2009er}%
  \BibitemOpen
  \bibfield  {author} {\bibinfo {author} {\bibfnamefont {A.~M.}\ \bibnamefont
  {L{\"a}uchli}}\ and\ \bibinfo {author} {\bibfnamefont {P.}~\bibnamefont
  {Werner}},\ }\href {\doibase 10.1103/PhysRevB.80.235117} {\bibfield
  {journal} {\bibinfo  {journal} {Physical Review B}\ }\textbf {\bibinfo
  {volume} {80}},\ \bibinfo {pages} {235117} (\bibinfo {year}
  {2009})}\BibitemShut {NoStop}%
\bibitem [{\citenamefont {Gorelov}\ \emph {et~al.}(2009)\citenamefont
  {Gorelov}, \citenamefont {Wehling}, \citenamefont {Rubtsov}, \citenamefont
  {Katsnelson},\ and\ \citenamefont {Lichtenstein}}]{Gorelov:2009bi}%
  \BibitemOpen
  \bibfield  {author} {\bibinfo {author} {\bibfnamefont {E.}~\bibnamefont
  {Gorelov}}, \bibinfo {author} {\bibfnamefont {T.~O.}\ \bibnamefont
  {Wehling}}, \bibinfo {author} {\bibfnamefont {A.~N.}\ \bibnamefont
  {Rubtsov}}, \bibinfo {author} {\bibfnamefont {M.~I.}\ \bibnamefont
  {Katsnelson}}, \ and\ \bibinfo {author} {\bibfnamefont {A.~I.}\ \bibnamefont
  {Lichtenstein}},\ }\href {\doibase 10.1103/PhysRevB.80.155132} {\bibfield
  {journal} {\bibinfo  {journal} {Physical Review B}\ }\textbf {\bibinfo
  {volume} {80}},\ \bibinfo {pages} {155132} (\bibinfo {year}
  {2009})}\BibitemShut {NoStop}%
\bibitem [{Note3()}]{Note3}%
  \BibitemOpen
  \bibinfo {note} {Orthogonal matrices are parameterized as $\protect
  \boldsymbol {U} = e^{\protect \boldsymbol {T}}$, where $\protect \boldsymbol
  {T}$ is an anti-symmetric real matrix ($T_{ij}=-T_{ji}$).~\cite
  {Culver:1966bj,Sherif:2008vu}}\BibitemShut {NoStop}%
\bibitem [{\citenamefont {Kirkpatrick}\ \emph {et~al.}(1983)\citenamefont
  {Kirkpatrick}, \citenamefont {Gelatt},\ and\ \citenamefont
  {Vecchi}}]{Kirkpatrick:1983zz}%
  \BibitemOpen
  \bibfield  {author} {\bibinfo {author} {\bibfnamefont {S.}~\bibnamefont
  {Kirkpatrick}}, \bibinfo {author} {\bibfnamefont {C.~D.}\ \bibnamefont
  {Gelatt}}, \ and\ \bibinfo {author} {\bibfnamefont {M.~P.}\ \bibnamefont
  {Vecchi}},\ }\href {\doibase 10.1126/science.220.4598.671} {\bibfield
  {journal} {\bibinfo  {journal} {Science}\ }\textbf {\bibinfo {volume}
  {220}},\ \bibinfo {pages} {671} (\bibinfo {year} {1983})}\BibitemShut
  {NoStop}%
\bibitem [{\citenamefont {{\v C}ern{\'{y}}}(1985)}]{Cerny:1985bg}%
  \BibitemOpen
  \bibfield  {author} {\bibinfo {author} {\bibfnamefont {V.}~\bibnamefont {{\v
  C}ern{\'{y}}}},\ }\href {\doibase 10.1007/BF00940812} {\bibfield  {journal}
  {\bibinfo  {journal} {Journal of optimization theory and applications}\
  }\textbf {\bibinfo {volume} {45}},\ \bibinfo {pages} {41} (\bibinfo {year}
  {1985})}\BibitemShut {NoStop}%
\bibitem [{\citenamefont {Bauer}\ \emph {et~al.}(2011)\citenamefont {Bauer},
  \citenamefont {Carr}, \citenamefont {Evertz}, \citenamefont {Feiguin},
  \citenamefont {Freire}, \citenamefont {Fuchs}, \citenamefont {Gamper},
  \citenamefont {Gukelberger}, \citenamefont {Gull},\ and\ \citenamefont
  {Guertler}}]{Bauer:2011tz}%
  \BibitemOpen
  \bibfield  {author} {\bibinfo {author} {\bibfnamefont {B.}~\bibnamefont
  {Bauer}}, \bibinfo {author} {\bibfnamefont {L.~D.}\ \bibnamefont {Carr}},
  \bibinfo {author} {\bibfnamefont {H.~G.}\ \bibnamefont {Evertz}}, \bibinfo
  {author} {\bibfnamefont {A.}~\bibnamefont {Feiguin}}, \bibinfo {author}
  {\bibfnamefont {J.}~\bibnamefont {Freire}}, \bibinfo {author} {\bibfnamefont
  {S.}~\bibnamefont {Fuchs}}, \bibinfo {author} {\bibfnamefont
  {L.}~\bibnamefont {Gamper}}, \bibinfo {author} {\bibfnamefont
  {J.}~\bibnamefont {Gukelberger}}, \bibinfo {author} {\bibfnamefont
  {E.}~\bibnamefont {Gull}}, \ and\ \bibinfo {author} {\bibfnamefont
  {S.}~\bibnamefont {Guertler}},\ }\href
  {http://iopscience.iop.org/1742-5468/2011/05/P05001} {\bibfield  {journal}
  {\bibinfo  {journal} {Journal of Statistical Mechanics: Theory and
  Experiment}\ }\textbf {\bibinfo {volume} {2011}},\ \bibinfo {pages} {P05001}
  (\bibinfo {year} {2011})}\BibitemShut {NoStop}%
\bibitem [{\citenamefont {Culver}(1966)}]{Culver:1966bj}%
  \BibitemOpen
  \bibfield  {author} {\bibinfo {author} {\bibfnamefont {W.~J.}\ \bibnamefont
  {Culver}},\ }in\ \href {\doibase 10.2307/2036109} {\emph {\bibinfo
  {booktitle} {Proceedings of the American Mathematical Society}}}\ (\bibinfo
  {year} {1966})\ p.\ \bibinfo {pages} {1146}\BibitemShut {NoStop}%
\bibitem [{\citenamefont {Sherif}\ and\ \citenamefont
  {Morsy}(2008)}]{Sherif:2008vu}%
  \BibitemOpen
  \bibfield  {author} {\bibinfo {author} {\bibfnamefont {N.}~\bibnamefont
  {Sherif}}\ and\ \bibinfo {author} {\bibfnamefont {E.}~\bibnamefont {Morsy}},\
  }\href
  {http://scuegypt.edu.eg/ar/wp-content/uploads/2013/02/morsyIJA1-4-2008.pdf}
  {\bibfield  {journal} {\bibinfo  {journal} {Int J Algebra}\ }\textbf
  {\bibinfo {volume} {2}},\ \bibinfo {pages} {131} (\bibinfo {year}
  {2008})}\BibitemShut {NoStop}%
\end{thebibliography}%

\appendix
\section{Single-particle rotation matrix}\label{sec:matrix}
We provide the single-particle rotation matrices used in the present study as text files.
%binary files in HDF5 format~\footnote{https://www.hdfgroup.org/HDF5/} as column matrices.
%A HDF5 binary file can be read easily, e.g., by using the h5py package for Python.

\subsection{Trimer model}
	%Rotation matrices for the trimer impurity model.
	The matrices for $t^\prime/t=0.5$ and $t^\prime/t=2$ are stored in ``rotmat-trimer-tratio0.5.txt'' and  ``rotmat-trimer-tratio2.0.txt'', respectively.
	The first, second, third columns are the indices $i$, $j$, and the entry $U_{ij}$, respectively.
	The graphs are numbered from $N_L=3$ in decreasing order and from No. 1 in increasing order as ``Graph 1, 2, $\cdots$".

\subsection{Tetramer model}
	%Rotation matrices for the tetramer impurity model.
	The matrices for $t^\prime/t=1.0$ and $t^\prime/t=0.5$ are stored in ``rotmat-tetramer-tratio1.0.txt'' and  ``rotmat-tetramer-tratio0.5.txt'', respectively.	
	The first, second, third columns are the indices $i$, $j$, and the entry $U_{ij}$, respectively.	
	The graphs are numbered from $N_L=6$ in decreasing order and from No. 1 in increasing order as ``Graph 1, 2, $\cdots$".

\section{Average sign computed for the trimer and tetramer models}\label{sec:sign-all}
\begin{figure*}
	\begin{tabular}{cc}
	\begin{minipage}{0.45\hsize}
		\centering\includegraphics[width=\textwidth,clip,type=pdf,ext=.pdf,read=.pdf]{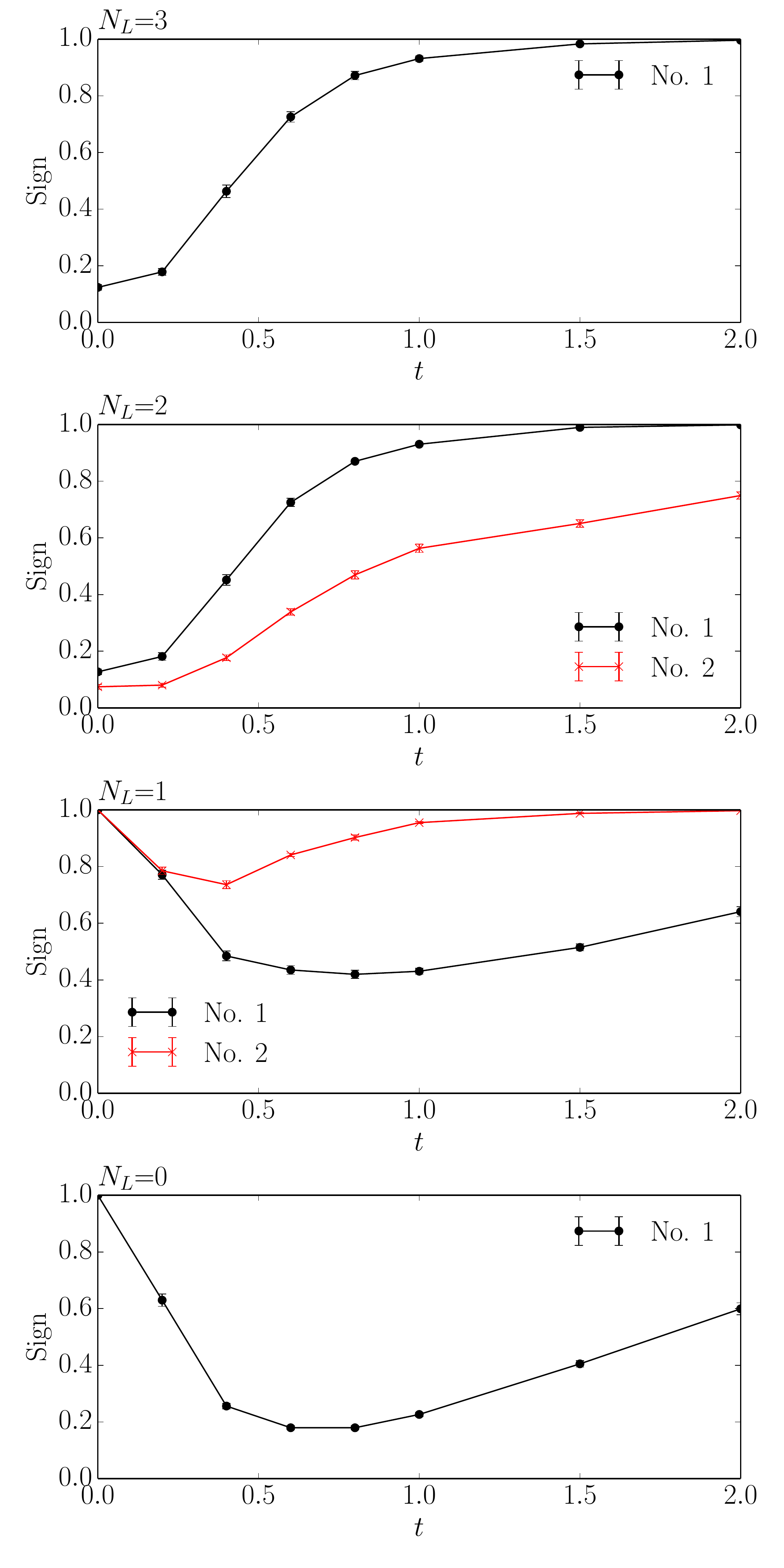}
	\end{minipage}
	&
	\begin{minipage}{0.45\hsize}
		\centering\includegraphics[width=\textwidth,clip,type=pdf,ext=.pdf,read=.pdf]{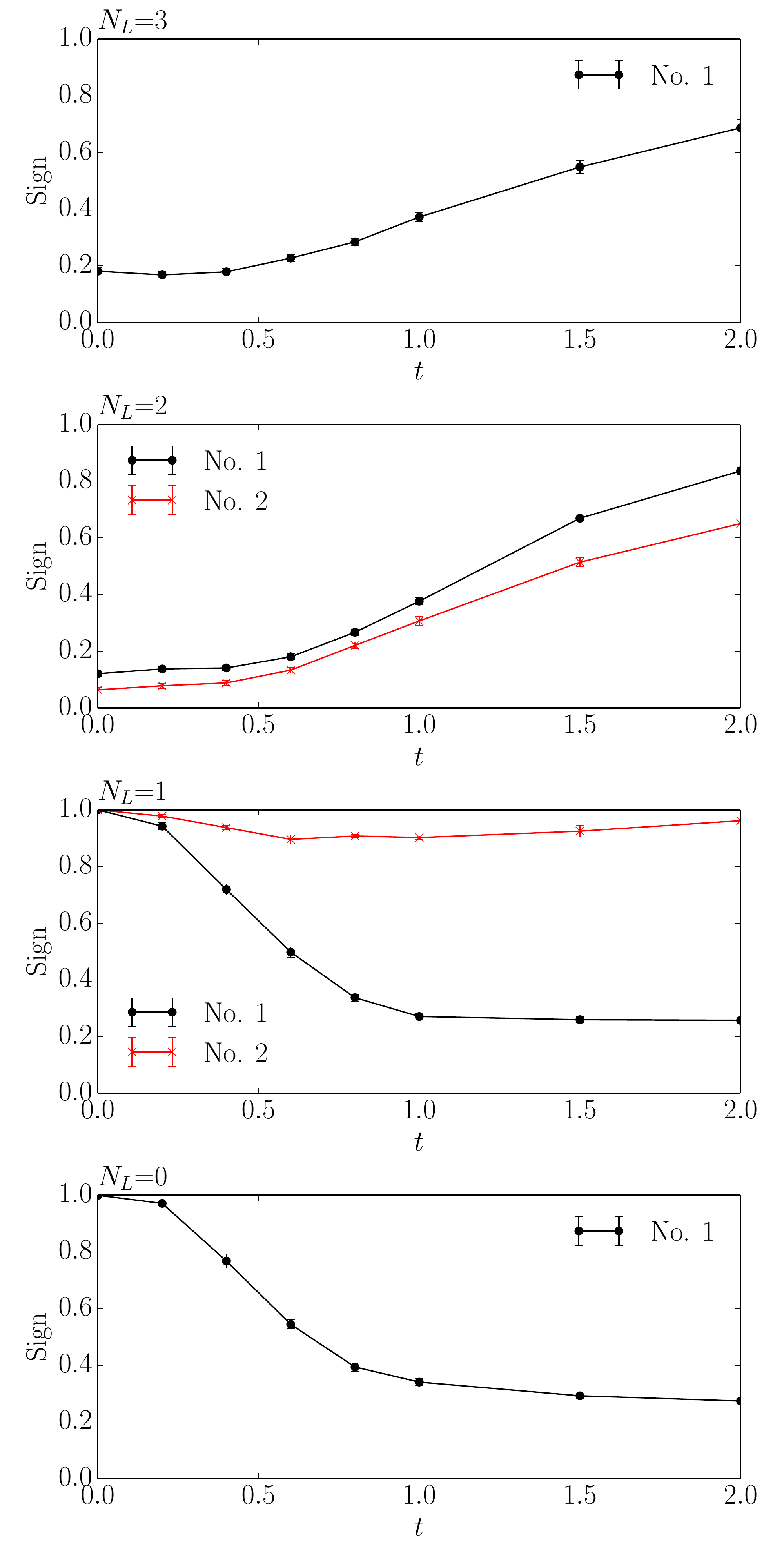}
	\end{minipage}	
    \end{tabular}
	\caption{
	(Color online) Average sign computed for the trimer model: (a) $t^\prime/t=2$ and (b) 0.5 at $\beta=50$.
	(Right panel) $t^\prime/t=2$ and (Left panel) $t^\prime/t=0.5$.
	We show the average sign computed for each basis.
	}
	\label{fig:trimer-sign-nosymm-all}
\end{figure*}

\begin{figure*}
	\begin{tabular}{cc}
		\begin{minipage}{0.45\hsize}
			\centering\includegraphics[width=\textwidth,clip,type=pdf,ext=.pdf,read=.pdf]{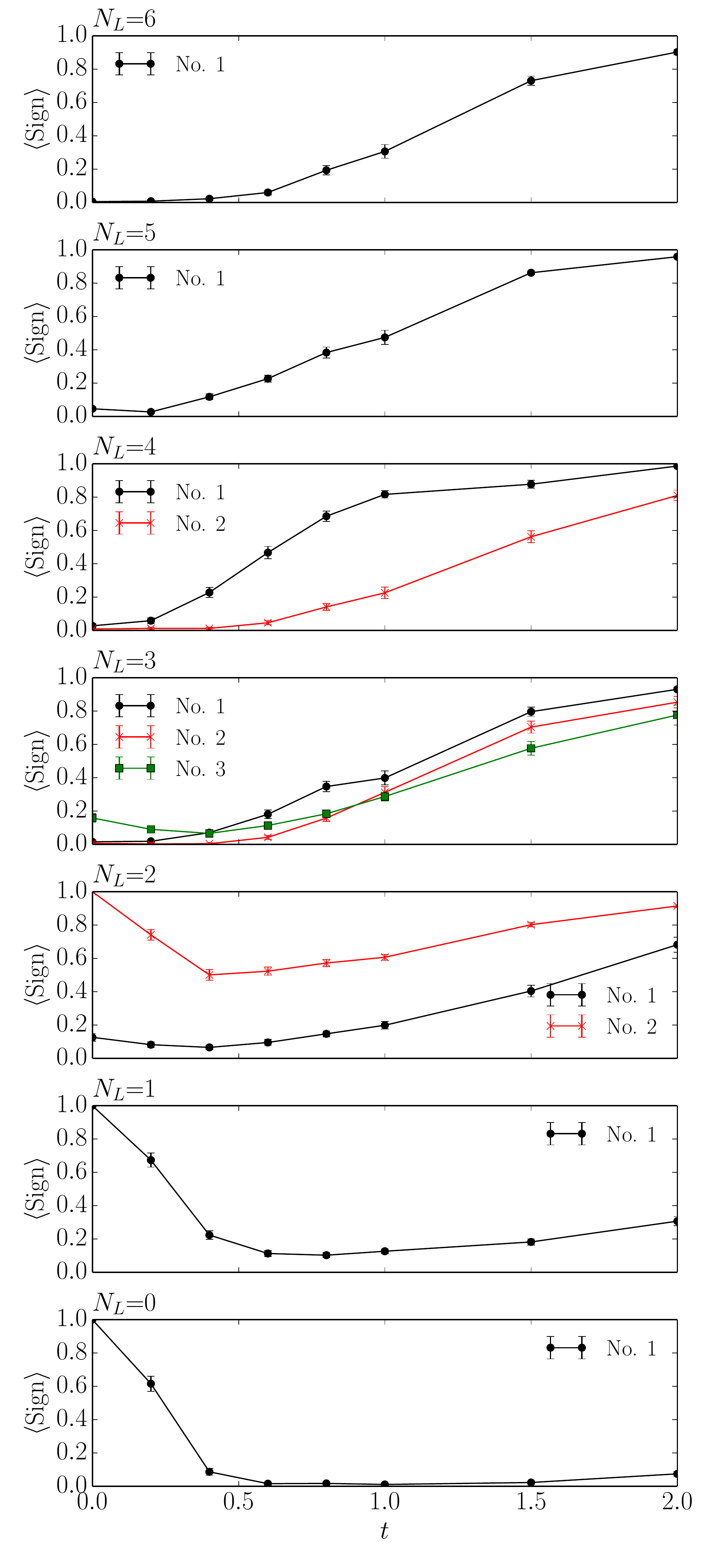}
		\end{minipage}
		&
		\begin{minipage}{0.45\hsize}
			\centering\includegraphics[width=\textwidth,clip,type=pdf,ext=.pdf,read=.pdf]{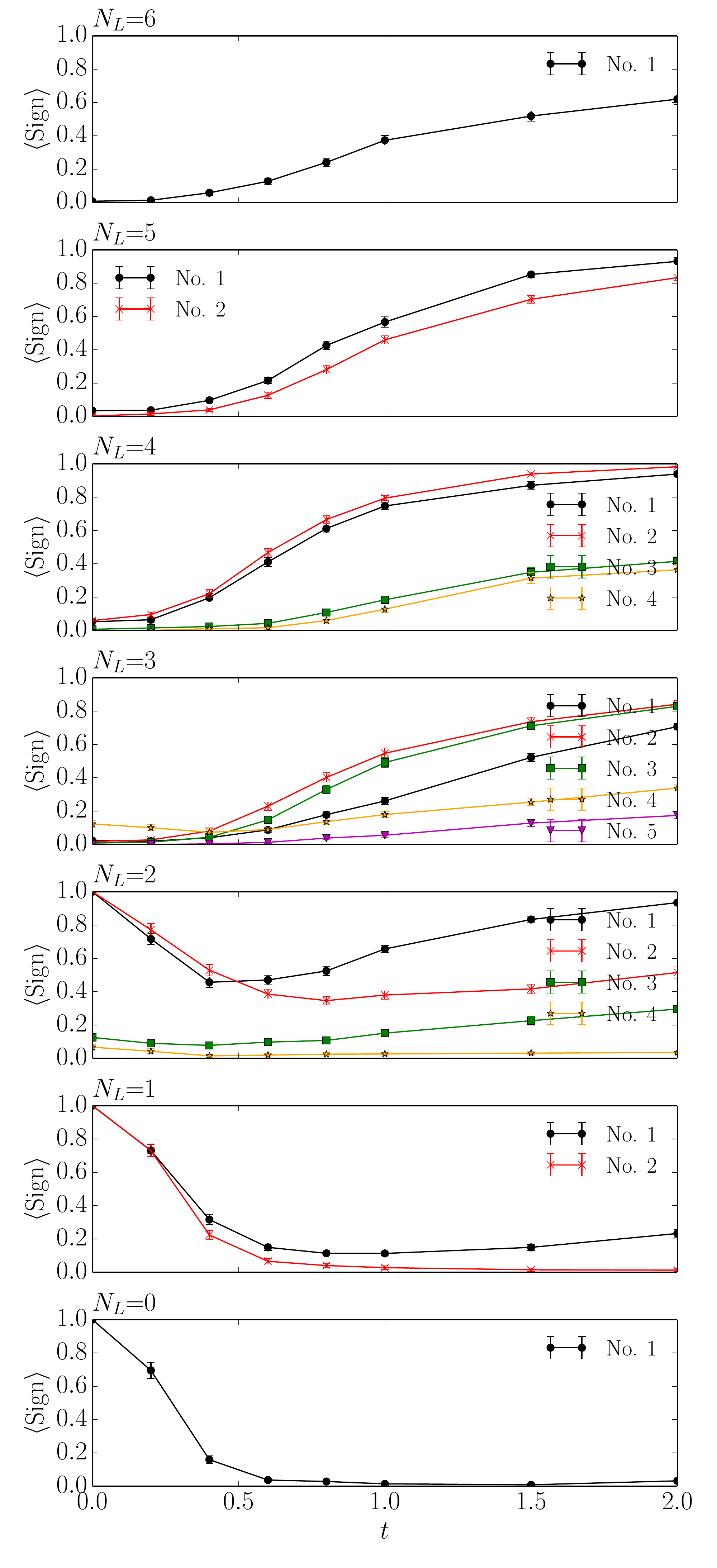}
		\end{minipage}	
	\end{tabular}
	\caption{
		(Color online) Average sign computed for the tetramer model at $\beta=50$:
		(Right panel) $t^\prime/t=1$ and (Left panel) $t^\prime/t=0.5$.
		We show the average sign computed for each basis. 
	}
	\label{fig:tetramer-sign-each-graph}
\end{figure*}

Figures~\ref{fig:trimer-sign-nosymm-all} and \ref{fig:tetramer-sign-each-graph} show the values of the average sign computed for all the graphs of the trimer and tetramer models, respectively.
See Figs.~\ref{fig:trimer} and \ref{fig:tetramer} for the structures of the graphs.

\section{Distribution function of the expansion order in continuous-time QMC}\label{sec:P}
%\tr{[PLEASE CHECK]}
We consider a quantum system whose Hamiltonian is given
\begin{eqnarray}
  H &=& H_0 + \lambda H_1~(\lambda\ge 0).
\end{eqnarray}
In a continuous-time QMC algorithm, we expand the partition function as
\begin{eqnarray}
 Z(\lambda)&=&\mathrm{Tr}[e^{-\beta H}]\nonumber\\
  &=& \sum_{n=0}^\infty \int_0^{\beta} d\tau_1 \cdots \int_{\tau_{n-1}}^\beta d \tau_n
  \lambda^n \nonumber\\
  && \times \mathrm{Tr}[(-1)^n e^{-(\beta-\tau_n)H_0} H_1 \cdots H_1 e^{-\tau_1 H_0}].
\end{eqnarray}
Practically, we decompose $H_1$ as
\begin{eqnarray}
 H_1 &=& \sum_\alpha H_{1\alpha},\label{eq:H1-decomposition}
\end{eqnarray}
depending on the basis to represent $H_1$.
Then, the partition function is given by
\begin{eqnarray}
 Z(\lambda) &=& \sum_{n=0}^\infty \int_0^{\beta} d\tau_1 \cdots \int_{\tau_{n-1}}^\beta d \tau_n
  \lambda^n w_{c_n},
\end{eqnarray}
where
\begin{eqnarray}
  \!\!\!\!\!\! w_{c_n} &\equiv & \mathrm{Tr}[(-1)^n e^{-(\beta-\tau_n)H_0} H_{1\alpha_1} \cdots H_{1\alpha_n} e^{-\tau_1 H_0}].
\end{eqnarray}

In a QMC simulation, we sample the partition function using $w (c_n)$ as MC weight.
In the following, we take $\lambda=1$.
The distribution function of $n$ is given by 
\begin{eqnarray}
    P(n) &=& \frac{\sum_{c_n} |w_{c_n}| }{\sum_{m=0}^\infty \sum_{c_m} |w_{c_m}|},~\label{eq:Pn}
\end{eqnarray}
where the summation $\sum_{c_n}$ is taken over all $n$-th order diagrams.
The average sign at each $n$ is defined as follows: 
\begin{eqnarray}
    \langle \mathrm{sign} \rangle_n &\equiv& \frac{\sum_{c_n} \mathrm{sign} (w_{c_n})|w_{c_n}| }{\sum_{c_n} |w_{c_n}|} = \frac{\sum_{c_n} w_{c_n} }{\sum_{c_n} |w_{c_n}|}.
\end{eqnarray}
$P(n)$ and $\langle \mathrm{sign} \rangle_n$ generally depend on the way of the decomposition of $H_1$ in Eq.~(\ref{eq:H1-decomposition}).
However, if $\langle \mathrm{sign} \rangle_n$ is independent of $n$,
$P(n)$ does not depend on the decomposition of $H_1$.
Assuming $\langle \mathrm{sign} \rangle_n = C$ ($C$ is a constant),
Eq.~(\ref{eq:Pn}) can be written as
\begin{eqnarray}
  P(n) &=& \frac{C^{-1}\sum_{c_n} w_{c_n} }{C^{-1}\sum_{m=0}^\infty \sum_{c_m} w_{c_m}}\nonumber\\
  &=& \frac{\sum_{c_n} w_{c_n} }{\sum_{m=0}^\infty \sum_{c_m} w_{c_m}}\nonumber\\
  %&=& \frac{\sum_{c_n} w_{c_n} }{Z}\nonumber\\
  &=& \frac{\frac{1}{n!}\frac{\partial^n Z(\lambda)}{\partial \lambda^n }|_{\lambda=0} }{Z(\lambda=1)}.~\label{eq:Pn-last}
\end{eqnarray}
One can easily see that the last line in Eq.~(\ref{eq:Pn-last}) depends only on how we decompose $H$ into $H_0$ and $H_1$, and hence does not depend on the basis in which $H_1$ is expressed.

%\begin{eqnarray}
%    \langle \mathrm{sign} \rangle_n &\equiv& \frac{\sum_{c_n} \mathrm{sign} (w_{c_n})|w_{c_n}| }{\sum_{m=0}^\infty \sum_{c_m} |w_{c_m}|}\nonumber\\
%    &=& \frac{\sum_{c_n} w_{c_n} }{\sum_{m=0}^\infty \sum_{c_m} |w_{c_m}|}.
%\end{eqnarray}
%Equation~(\label{eq:Pn}) reads

\section{Coulomb interaction for bonding and anti-bonding orbitals}\label{sec:coulomb-bonding}
Here, we derive the Coulomb interaction for bonding and anti-bonding orbitals.
We consider a two-site model with the onsite Coulomb interaction $U$ whose Hamiltonian is given in the site basis as follows:
\begin{eqnarray}
	\mathcal{H} &=& U\sum_{i=1}^2 n_{i\uparrow}n_{i\downarrow},~\label{eq:two-site}
\end{eqnarray}
where $n_{i\sigma} = c^\dagger_{i\sigma} c_{i\sigma}$.

This Hamiltonian can be expressed in the form of a Slater-Kanamori interaction as
\begin{eqnarray}
	&&\mathcal{H}=\frac{1}{2}\sum_{\alpha\beta\alpha^\prime\beta^\prime}^2\sum_{\sigma\sigma^\prime}U_{\alpha\beta\alpha^\prime\beta^\prime}c^\dagger_{\alpha\sigma}c^\dagger_{\beta\sigma^\prime}c_{\beta^\prime\sigma^\prime}c_{\alpha^\prime\sigma},\label{eq:dimer-SK}
\end{eqnarray}
with the intra-orbital repulsion $U_{\alpha\alpha\alpha\alpha}=U$, the inter-orbital repulsion $U_{\alpha\beta\alpha\beta}=U^\prime=0$, the exchange interaction $U_{\alpha\beta\beta\alpha}=J_\mathrm{H}=0$, and the pair hopping $U_{\alpha\alpha\beta\beta}=J_\mathrm{H}^\prime=0$ ($\alpha\neq \beta$).

We now consider the following basis rotation ($0\le \theta\le \pi/4$):
\begin{eqnarray}
	d_a &=& \sum_b V_{ba}^* c_b,\\
	d^\dagger_a &=& \sum_b V_{ba} c^\dagger_b,\\
	\boldsymbol{V} &=& 
	\left(
	\begin{array}{cc}
		\cos\theta & -\sin\theta\\ 
		\sin\theta & \cos\theta\\
	\end{array}
	\right).~\label{eq:dimer-rotation}
\end{eqnarray}
The site basis is given by $\theta=0$, while $\theta=\pi/4$ corresponds to a dimer basis.

Substituting Eq.~(\ref{eq:dimer-rotation}) into Eq.~(\ref{eq:dimer-SK}),
one obtains
\begin{eqnarray}
	\mathcal{H} &=&\frac{1}{2}\sum_{\alpha\beta\alpha^\prime\beta^\prime}\sum_{\sigma\sigma^\prime}U_{\alpha\beta\alpha^\prime\beta^\prime}c^\dagger_{\alpha\sigma}c^\dagger_{\beta\sigma^\prime}c_{\beta^\prime\sigma^\prime}c_{\alpha^\prime\sigma}\nonumber\\
	&=&\frac{1}{2}
	\sum_{ijkl}
	\sum_{\sigma\sigma^\prime}
	\bar{U}_{ijkl}
	d^\dagger_{i\sigma}
	d^\dagger_{j\sigma^\prime}
	d_{l\sigma^\prime}
	d_{k\sigma}.
\end{eqnarray}
The Coulomb matrix in the rotated basis reads
\begin{eqnarray}
	\bar{U}_{ijkl} &=& \sum_{\alpha\beta\alpha^\prime\beta^\prime} U_{\alpha\beta\alpha^\prime\beta^\prime} V^*_{\alpha i}
	V^*_{\beta j}
	V_{\alpha^\prime k}
	V_{\beta^\prime l},
\end{eqnarray}
where $i$, $j$, $k$, $l$ are the index of orbitals in the rotated basis.
The nonzero elements in $\bar{U}_{ijkl}$ are 
\begin{eqnarray}
	\bar{U}\equiv \bar{U}_{iiii}= (\cos^4\theta+\sin^4\theta)U,\\
	\bar{U}^\prime \equiv \bar{U}_{ijij}= (2\cos^2 \theta \sin^2 \theta)U,\\
	\bar{J}_\mathrm{H} \equiv \bar{U}_{ijji}= (2\cos^2 \theta \sin^2 \theta)U,\\
	\bar{J}_\mathrm{H}^\prime \equiv \bar{U}_{iijj}=(2\cos^2 \theta \sin^2 \theta)U,
\end{eqnarray}
where $i\neq j$.
For the dimer basis ($\theta=\pi/4$),
one obtains $\bar{U}=\bar{U}^\prime=\bar{J}_\mathrm{H}=\bar{J}_\mathrm{H}^\prime=U/2$.

\section{Improved double vertex update}\label{sec:double-vertex-update}
\begin{figure}
	\includegraphics[width=.49\textwidth,clip]{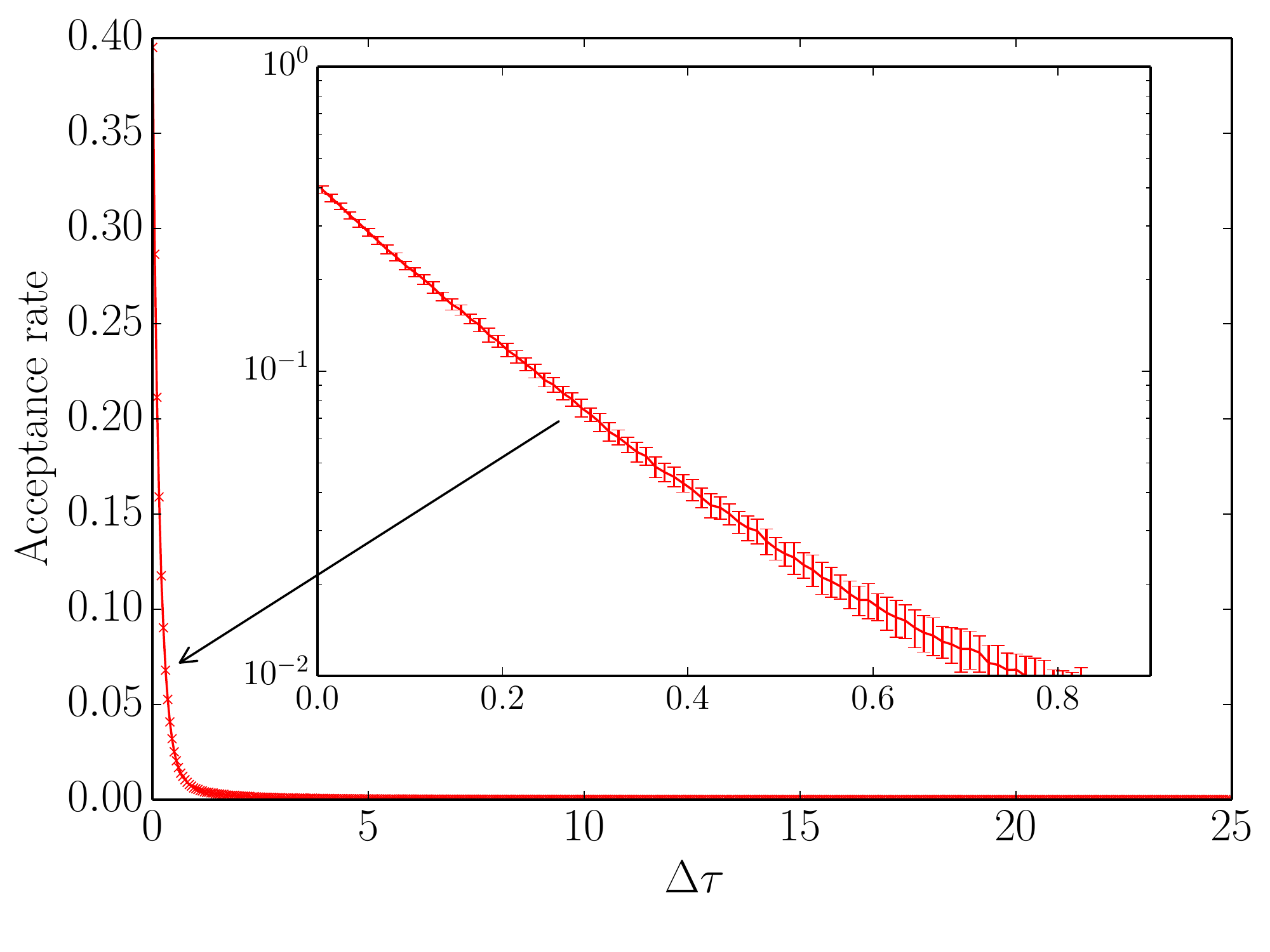}
	\caption{
	(Color online)
    Acceptance rate for the insertion of a double vertex as a function of the time difference between two vertices ($\Delta \tau$) computed for the tetramer model at $U=8$, $\beta=50$, and $t=t^\prime=1$.
    The data are symmetrized about $\beta/2$ and are averaged over spin-flip interactions and pair-hopping interactions.
    }
	\label{fig:acc_rate}
\end{figure}

In Ref.~\onlinecite{Nomura:2014by}, an efficient CT-INT algorithm, the so-called ``double vertex update", has been proposed. 
The algorithm has been developed to efficiently deal with non-density type interactions (spin-flip and pair-hopping interactions). 
In this update, one inserts or removes a pair of two vertices corresponding to spin-flip or pair-hopping interactions simultaneously. 
In the present study, we employ this scheme to deal with the non-density-type interactions generated by the basis transformation.
However, we found that the acceptance ratio becomes very low at large $\beta$ if we insert a pair of vertices at two imaginary times randomly picked from the whole interval $[0, \beta)$ as in Ref.~\onlinecite{Nomura:2014by}.
This is because the acceptance ratio drops exponentially with the time difference between the two vertices (see Fig.~\ref{fig:acc_rate}).

To overcome this problem, we improve the scheme by increasing the proposal ratio for double-vertex insertions with short time differences.  
Let us give a detailed explanation by taking the update of two spin-flip interactions as an example.
The scheme for the pair-hopping interactions can be formulated in an analogous way.
For the spin-flip interactions, there are two different types of vertices,
$J d^{\dagger}_{\alpha \uparrow} d^{\dagger}_{\beta \downarrow} d_{\alpha \downarrow} d_{\beta \uparrow}$  and $J d^{\dagger}_{\beta \uparrow} d^{\dagger}_{\alpha \downarrow} d_{\beta \downarrow} d_{\alpha \uparrow}$,
which appear in pairs during the MC simulation.
That is, the perturbation orders of these two types of vertices are always the same, which we denote by $m_{\rm s}$. 

In the improved scheme, for the insertion, we pick an imaginary time $\tau_1$ randomly in $[0, \beta]$ for one type of the spin-flip interactions. 
Then, we propose the second imaginary time $\tau_2^\prime$ for the other type of the spin-flip interaction with a probability proportional to $p(\Delta \tau)$ given by 
\begin{eqnarray}
\label{Eq_ptau}
 p(\Delta \tau) = e^{- a\Delta \tau}  +  e^{-a(\beta - \Delta \tau)}   + b
\end{eqnarray} 
with $\Delta \tau = |\tau_2^\prime - \tau_1|$. 
The positive constants $a$ and $b$ should be optimized case by case.
For example, the optimal value $a$ may depend sensitively on the interaction strength.
On the other hand, $b$ can generally be chosen small, $b\sim 10^{-3}$--$10^{-2}$. 
$ p(\Delta \tau)$ is defined in the region $\Delta \tau \in [ 0 , \beta ]$.
One can see that $p(\Delta \tau)$ has a larger value around $\Delta \tau \sim 0$ and $\Delta \tau \sim \beta$, 
i.e., we propose the insertion of two vertices with a short time difference more frequently.
The proposal weight for the insertion of two vertices with time difference $\Delta \tau$ is 
\begin{eqnarray}
P_{\rm add} (m_{\rm s} \rightarrow m_{\rm s}\!+\!1) =   \left( \frac{d\tau}{\beta} \right)^2  \! \frac{ p(\Delta \tau) } {  \frac{2}{a \beta}  \left (1 - {\rm exp}^{-a \beta}  \right ) + b  }, \hspace{5mm}
\end{eqnarray}
where the factor $\bigl [ \frac{2}{a \beta}  \left (1 - {\rm exp}^{-a \beta}  \right ) + b \bigr ]$  originates from the integral $\frac{1}{\beta} \int_{0}^{\beta} p(\tau) d\tau$. 

In the removal update, we pick a pair of spin-flip vertices with a probability proportional to 
$p(\Delta \tau)$ with $\Delta \tau$ being the time difference of the two vertices. 
Then the proposal weight for the removal of the pair of the $I$th spin-flip vertex of type 1 at $\tau_I$ and $J$th spin-flip vertex of type 2 at $\tau'_J$ is 
\begin{eqnarray}
P_{\rm rm} (m_{\rm s}\!+\!1 \rightarrow m_{\rm s}) =  p\bigl( | \tau_I - \tau'_J | \bigr ) /  F_{m_{\rm s}}, 
\end{eqnarray}
where $F_{m_{\rm S}}$ is calculated as 
\begin{eqnarray}
\label{Eq.F_m}
F_{m_{\rm s} } = \sum_{i,j = 1}^{m_{\rm s} }   p\bigl( | \tau_i - \tau'_j | \bigr ),
\end{eqnarray}
with $\{ \tau_i \}$ ($\{ \tau'_j\}$) being the imaginary times for the existing spin-flip vertices of type 1 (type 2).

Then the acceptance rate for the improved double vertex update becomes 
\begin{eqnarray}
P(c_n \! \rightarrow \! c_{n+2}) = {\rm min} \left( X 
\prod_{\sigma}  \biggl |  \frac{ {\rm det} A_{\sigma}(c_{n+2})}{{\rm det} A_{\sigma}(c_n)} \biggr | , 1   \right)
\end{eqnarray}
for the insertion update
 and 
\begin{eqnarray}
P(c_{n+2} \! \rightarrow \! c_{n}) = {\rm min} \!  \left( \frac{1}{X}
\prod_{\sigma}   \biggl |   \frac{{\rm det} A_{\sigma}(c_n)}{ {\rm det} A_{\sigma}(c_{n+2})}  \biggl |  , 1   \right)
\end{eqnarray} 
 for the removal update. 
 Here, $\det A$ determines the weight of the configuration in the CT-INT scheme, which originates from the Wick's decomposition of the operator series.~\cite{Gull:2011jda} 
 $n$ is the number of vertices in the configuration including both the density-type and non-density-type interactions.  
 The factor $X$ is given by 
 \begin{eqnarray}
 X &=& (J d\tau )^2 \frac{P_{\rm rm} (m_{\rm s}\!+\!1 \rightarrow m_{\rm s}) }{P_{\rm add}  (m_{\rm s} \rightarrow m_{\rm s}\!+\!1) }  \nonumber \\ 
 &=&   \frac{  (\beta J)^2  } { F_{m_{\rm s} + 1} }   \left[  \frac{2}{a \beta}  \left (1 - {\rm exp}^{-a \beta}  \right ) + b \   \right ].
 \label{Eq.expression_X}
 \end{eqnarray}

For the tetramer, without the improvement presented here, the acceptance ratio for the double vertex update is as small as about 0.003 for $\beta = 50$, $U=8$, and $t=t^\prime=1$.
We obtained $a \sim 5.8$ and $b \sim 0.002$ by fitting the acceptance rate in Fig.~\ref{fig:acc_rate} using the expression for $p(\tau)$ in Eq.~(\ref{Eq_ptau}) for $0 \le \tau \le 15$.
We found that with these parameters, the acceptance rate is dramatically improved from $\simeq$ 0.003 to $\simeq$ 0.1.

\end{document}